%% file: SimpleSets.tex
\documentclass[journal]{vgtc}                     

\onlineid{0}

\vgtccategory{Research}

\title{SimpleSets: Capturing Categorical Point Patterns\texorpdfstring{\\}{} with Simple Shapes}

\author{%
  \authororcid{Steven van den Broek}{0009-0005-6677-3916},
  \authororcid{Wouter Meulemans}{0000-0002-4978-3400}, and 
  \authororcid{Bettina Speckmann}{0000-0002-8514-7858}
}

\authorfooter{
  	TU Eindhoven, the Netherlands. E-mail:\newline [s.w.v.d.broek, w.meulemans, b.speckmann]@tue.nl.
   
}

\abstract{%
Points of interest on a map such as restaurants, hotels, or subway stations, give rise to categorical point data: data that have a fixed location and one or more categorical attributes.
    Consequently, recent years have seen various set visualization approaches that visually connect points of the same category to support users in understanding the spatial distribution of categories.
    Existing methods use complex and often highly irregular shapes to connect points of the same category, leading to high cognitive load for the user.
    In this paper we introduce SimpleSets, which uses simple shapes to enclose categorical point patterns, thereby providing a clean overview of the data distribution.
    SimpleSets is designed to visualize sets of points with a single categorical attribute; as a result, the point patterns enclosed by SimpleSets form a partition of the data.
    We give formal definitions of point patterns that correspond to simple shapes and describe an algorithm that partitions categorical points into few such patterns.
    Our second contribution is a rendering algorithm that transforms a given partition into a clean set of shapes resulting in an aesthetically pleasing set visualization.
    Our algorithm pays particular attention to resolving intersections between nearby shapes in a consistent manner.
    We compare SimpleSets to the state-of-the-art set visualizations using standard datasets from the literature.

}


\teaser{%
    \centering\includegraphics[width=0.8\linewidth]{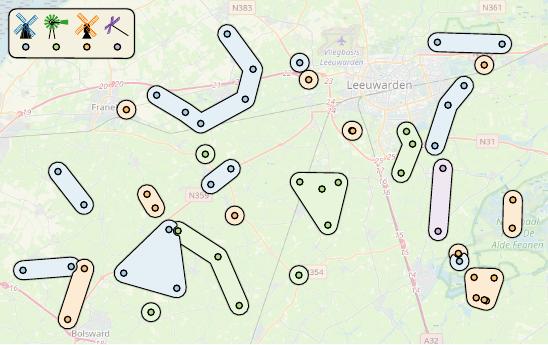}
  \caption{A SimpleSets visualization of mills around Leeuwarden, The Netherlands. The mill types are: angular mill (blue); vertical wind engine (green); spider head mill (orange); and tjasker (purple). Data by \href{https://molendatabase.nl}{molendatabase.nl} with permission, map from \href{https://www.openstreetmap.org/copyright}{openstreetmap.org}.}%
\label{fig:teaser}%
}

\graphicspath{{figs/}{figures/}{pictures/}{images/}{./}} 

\usepackage{subcaption}
\usepackage{amsmath}
\usepackage[ruled,vlined]{algorithm2e}
\SetKwBlock{Repeat}{repeat}{}
\DontPrintSemicolon
\SetArgSty{textnormal} 

\SetCommentSty{mycommfont}
\SetKwInput{Input}{Input}
\SetKwInput{Output}{Output}

\usepackage{bibunits}
\defaultbibliographystyle{abbrv-doi-hyperref-last-name-sorting}
\defaultbibliography{bibliography.bib}

\usepackage{tabu}
\usepackage{booktabs}
\usepackage{multirow}

\usepackage{enumitem}

\usepackage[svgnames]{xcolor}
\usepackage{pgfplots}
\pgfplotsset{compat=1.9} 

\definecolor{CBblue}{rgb}{0.121, 0.47, 0.705}
\definecolor{CBgreen}{rgb}{0.2, 0.627, 0.172}
\definecolor{CBorange}{rgb}{1, 0.498, 0}
\definecolor{CBpurple}{rgb}{0.415, 0.239, 0.603}
\definecolor{CBred}{rgb}{0.89, 0.102, 0.109}

\newcommand{\cparagraph}[1]{\par\vspace{2mm}\noindent\textbf{#1}}

\usepackage{newtxmath}
\usepackage{stfloats}
\usepackage{flushend}

\input{fix-bibunit-hyperref}

\begin{document}

\makeatletter 
\begin{bibunit}
\input{SimpleSets-body}

\putbib
\end{bibunit}

\clearpage
\appendix

\renewcommand*{\HyperDestNameFilter}[1]{#1-appendix}

\input{SimpleSets-appendix-body}
\makeatother

\end{document}

%% file: fix-bibunit-hyperref.tex
 \makeatletter
\def\hyper@natlinkstart#1{%
  \Hy@backout{#1}%
  \hyper@linkstart{cite}{cite.\@bibunitname.#1}%
  \def\hyper@nat@current{#1}%
}

\def\hyper@natlinkbreak#1#2{%
  \hyper@linkend#1\hyper@linkstart{cite}{cite.\@bibunitname.#2}%
}

\def\hyper@natanchorstart#1{%
  \hyper@anchorstart{cite.\@bibunitname.#1}%
}

\def\bibcite#1#2{%
  \@newl@bel{b}{#1}{\hyper@@link[cite]{}{cite.\@bibunitname.#1}{#2}}%
}%

\def\@lbibitem[#1]#2{%
  \@skiphyperreftrue
  \H@item[\hyper@anchorstart{cite.\@bibunitname.#2}%
  \@BIBLABEL{#1}\hyper@anchorend\hfill]%
  \@skiphyperreffalse
  \if@filesw
    \begingroup
      \let\protect\noexpand
      \immediate\write\@auxout{%
        \string\bibcite{#2}{#1}%
      }%
    \endgroup
  \fi
  \ignorespaces
}%

\def\@bibitem#1{%
  \@skiphyperreftrue\H@item\@skiphyperreffalse
  \hyper@anchorstart{cite.\@bibunitname.#1}\relax\hyper@anchorend
  \if@filesw
    \begingroup
      \let\protect\noexpand
      \immediate\write\@auxout{%
        \string\bibcite{#1}{\the\value{\@listctr}}%
      }%
    \endgroup
  \fi
  \ignorespaces
}%

\def\@citex[#1]#2{%
  \let\@citea\@empty
  \@cite{%
    \@for\@citeb:=#2\do{%
      \@citea
      \def\@citea{,\penalty\@m\ }%
      \edef\@citeb{\expandafter\@firstofone\@citeb}%
      \if@filesw
        \immediate\write\@auxout{\string\citation{\@citeb}}%
      \fi
      \@ifundefined{b@\@citeb}{%
        \mbox{\reset@font\bfseries ?}%
        \G@refundefinedtrue
        \@latex@warning{%
          Citation `\@citeb' on page \thepage \space undefined%
        }%
      }{%
        \hyper@natlinkstart{\@citeb}%
            \hbox{\csname b@\@citeb\endcsname}%
        \hyper@natlinkend%
      }%
    }%
  }{#1}%
}%

\makeatother

%% file: SimpleSets-body.tex
\firstsection{Introduction}

\maketitle

Categorical point data are a common data type that consist of a location in the plane (or possibly in a higher dimension) labelled with a set of one or more categorical attributes. 
This data type captures, for example, points of interest on a map, such as restaurants or hotels, or the nodes of an embedded graph such as a social network.
In recent years, various approaches have been developed that visually connect the points that belong to the same category. 
Such connecting shapes aid the user in identifying patterns and understanding the data distribution. 
However, the majority of existing methods use one single shape to connect all points in the same category.
Since these points can be scattered throughout the plane, the shapes that represent the categories are often complex and may intersect even when the categories have no shared points, leading to a high cognitive load for the user.

We present SimpleSets, which uses simple enclosing shapes to capture patterns in point data with a single categorical attribute (Figure~\ref{fig:teaser}).
SimpleSets provides a clean overview of the data distribution, highlights natural spatial patterns, and covers little space in addition to the points.
The patterns constructed by SimpleSets depend on one intuitive parameter which allows the user to group points that are further apart or closer together in patterns that cover more or less space, as desired.

\begin{figure*}[tb]
    \begin{subfigure}{0.325\textwidth}
        \centering
        \frame{\includegraphics[width=\textwidth, page=1]{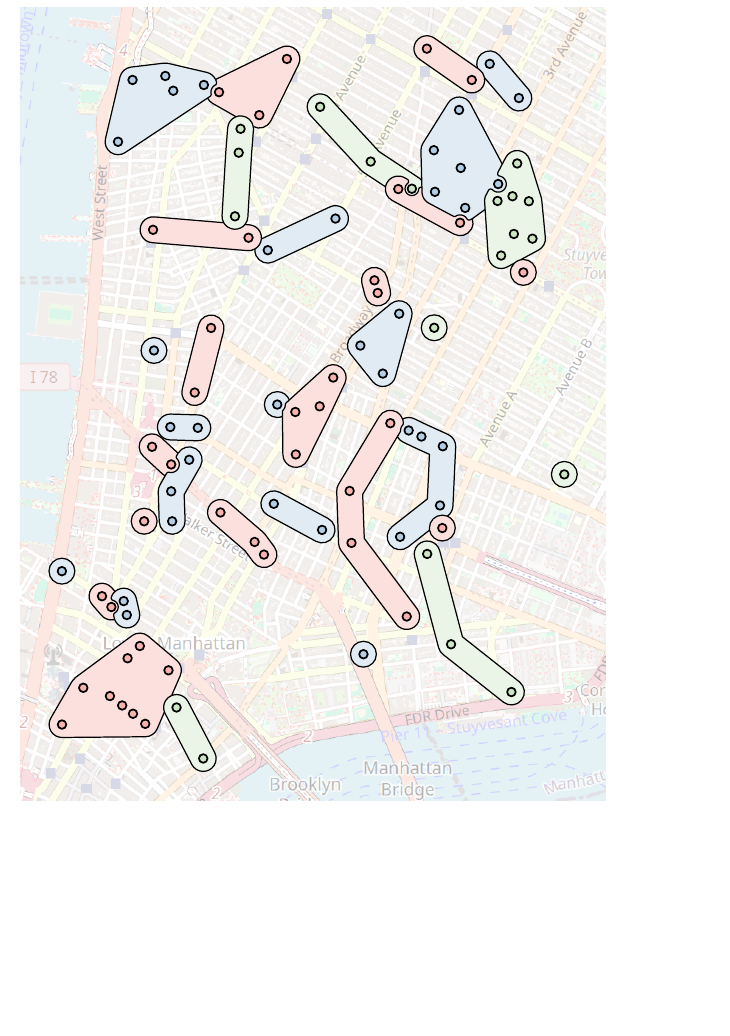}}
        \caption{SimpleSets}
        \label{fig:nyc:SimpleSets}
    \end{subfigure}
    \hfill
    \begin{subfigure}{0.325\textwidth}
        \centering
        \frame{\includegraphics[width=\textwidth, page=2]{nyc-comparison.pdf}}
        \caption{VPF*14~\cite{inverse-distance} (VPF*14+)}
        \label{fig:nyc:Vihrovs}
    \end{subfigure}
    \hfill
    \begin{subfigure}{0.325\textwidth}
        \centering
        \frame{\includegraphics[width=\textwidth, page=3]{nyc-comparison.pdf}}
        \caption{ClusterSets~\cite{ClusterSets} (hull-based drawing)}
        \label{fig:nyc:ClusterSets}
    \end{subfigure}

    \vspace{3mm}

    \begin{subfigure}{0.241\textwidth}
        \centering
        \frame{\includegraphics[width=\textwidth, page=4]{nyc-comparison.pdf}}
        \caption{Bubble Sets~\cite{BubbleSets}}
        \label{fig:nyc:BubbleSets}
    \end{subfigure}
    \hfill
    \begin{subfigure}{0.241\textwidth}
        \centering
        \frame{\includegraphics[width=\textwidth, page=5]{nyc-comparison.pdf}}
        \caption{LineSets~\cite{LineSets}}
        \label{fig:nyc:LineSets}
    \end{subfigure}
    \hfill
    \begin{subfigure}{0.241\textwidth}
        \centering
        \frame{\includegraphics[width=\textwidth, page=6]{nyc-comparison.pdf}}
        \caption{KelpFusion~\cite{KelpFusion}}
        \label{fig:nyc:KelpFusion}
    \end{subfigure}
    \hfill
    \begin{subfigure}{0.241\textwidth}
        \centering
        \frame{\includegraphics[width=\textwidth, page=7]{nyc-comparison.pdf}}
        \caption{MapSets~\cite{MapSets}}
        \label{fig:nyc:MapSets}
    \end{subfigure}

    \caption{
        A comparison of visualizations on a common benchmark dataset originating from the paper that introduced Bubble Sets~\cite{BubbleSets}. 
        The points show hotels (blue), subway entrances (red), and medical clinics (green) in lower Manhattan. 
    }
    \label{fig:nyc}
\end{figure*}

\cparagraph{Organization.}
We begin in Section~\ref{sec:related} with an extensive overview of related work.
SimpleSets consists of two phases: first, the point data is partitioned into spatial patterns, and then these patterns are drawn as enclosing shapes. 
In Section~\ref{sec:overview} we first discuss the overarching design decisions that underlie SimpleSets. Then, in Section~\ref{sec:constructing-partition} we describe how to partition the input into two types of simple patterns: islands and banks.
Intuitively, islands are convex clusters of points and banks are sequences of points that do not bend too much.
In Section~\ref{sec:drawing-partition}, we describe our drawing algorithm in detail; note that we can render any set of spatial point patterns, not only the ones defined by SimpleSets. In the supplementary material we show how to render other patterns from the state-of-the-art with our pipeline.
Finally, in Section~\ref{sec:discussion} we discuss SimpleSets visualizations for several datasets from the literature and compare to the state-of-the-art. We implemented our algorithm; our code is open-source to facilitate re-use and reproducibility and is available at \href{https://github.com/tue-alga/SimpleSets}{github.com/tue-alga/SimpleSets} and \href{https://doi.org/10.5281/zenodo.12784670}{doi:10.5281/zenodo.12784670}. Our implementation also includes code for the two methods most related to ours: VPF*14~\cite{inverse-distance} and ClusterSets~\cite{ClusterSets}; we re-implemented both methods since the original code was not or only partially available.
We close with a discussion of possible extensions and future work.

\section{Related Work}\label{sec:related}
The problem we address in this paper is a set visualization problem with points at predefined positions as elements and categorical attributes as sets.
Set visualization in general is a well-studied problem, and a range of visualization techniques have been used to visualize sets, from the well-known Euler and Venn diagrams, to matrix-based techniques.
We refer to the survey by Alsallakh et al.~\cite{set-survey} for an overview of set visualization research before 2016, and to Paetzold et al.~\cite{DBLP:journals/cgf/PaetzoldKSXSD23} for more recent results.
Convex shapes often feature in set visualizations such as Euler diagrams~\cite{untangling-euler-diagrams, DBLP:journals/cgf/PaetzoldKSXSD23} or node-link diagram overlays~\cite{social-networks-convex2, social-networks-convex1}, as Gestalt theory indicates that they provide a good sense of grouping~\cite{gestalt}.
This motivates the use of convex islands in our visualization.

In the remainder we discuss work on set visualizations that use predefined positions.
Dinkla et al.~\cite{kelp-diagrams} provide a detailed analysis of this problem. 
They list various tasks that a user may want to perform using the set visualization, constraints that any visualization should satisfy, and criteria that make shapes effective for set visualization. 
The most important criteria are:
\begin{enumerate}[noitemsep, label=C\arabic*, ref=C\arabic*]
    \item Low cognitive load.\label{c:cognitive}
    \item Strong continuation of shapes that depict the same set.\label{c:continuation}
    \item Little obfuscation of a possible underlying visualization.\label{c:obfuscation}
    \item Little distortion of point position and density.\label{c:distortion}
\end{enumerate}

\begin{figure}[t]
    \begin{subfigure}{0.25\columnwidth}
        \centering
        \includegraphics[page=2]{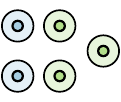}
        \caption{}
    \end{subfigure}
    \hfill
    \begin{subfigure}{0.25\columnwidth}
        \centering
        \includegraphics[page=4]{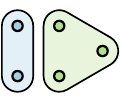}
        \caption{}
    \end{subfigure}
    \hfill
    \begin{subfigure}{0.27\columnwidth}
        \centering
        \includegraphics[page=6]{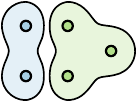}
        \caption{}
    \end{subfigure}
    \hfill
    \begin{subfigure}{0.18\columnwidth}
        \centering
        \includegraphics[page=8]{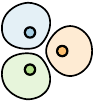}
        \caption{}
    \end{subfigure}
    \captionsetup{skip=1mm}
    \caption{Figure (c) uses arguably more complex shapes than (b) resulting in higher cognitive load (\ref{c:cognitive}).
    Figures (b) and (c), compared to (a), use fewer and larger shapes resulting in better continuation (\ref{c:continuation}) but more obfuscation (\ref{c:obfuscation}) and distortion (\ref{c:distortion}).  Figure (d) distorts point position compared to (a); the expected point position is at the centroid of a shape.}
    \label{fig:criteria}
\end{figure}

These criteria are illustrated in Figure~\ref{fig:criteria}. Most existing approaches focus on the second criterion and use only a single connecting shape which tends to negatively impact the other criteria.
SimpleSets instead relies on color to visually connect disjoint shapes that depict the same set, and explicitly connects points only if they form a simple spatial pattern. This allows SimpleSets to do well on the other criteria.

\begin{figure*}[b]
    \centering
    \includegraphics[page=5]{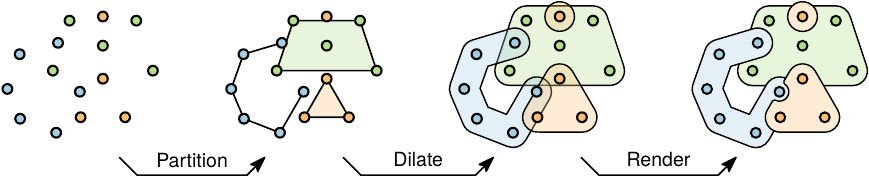}
    \caption{SimpleSets pipeline. From left to right: data points; partition into patterns; dilated patterns; final enclosing shapes.}
    \label{fig:overview}
\end{figure*}

In their survey, Alsallakh et al.~\cite{set-survey} identify two types of visual representations of sets: line-based and region-based overlays.
We refine the region-based category by distinguishing hull-based and Voronoi-based visualizations.
The approaches differ in the size of areas that are colored.
Line-based approaches color only a small area around points and around links that connect points.
Hull-based approaches color larger regions, and Voronoi-based approaches color the majority of the area being visualized.
With the three categories as a guide, we give an overview of related work. Figure~\ref{fig:nyc} shows most visualizations we discuss in the following on a commonly used benchmark dataset; we also include SimpleSets. In Section~\ref{sec:discussion} we discuss the visualizations most similar to SimpleSets in more detail and in comparison to SimpleSets.

\cparagraph{Hull-based.} 
Byelas and Telea~\cite{AOI} use a convex hull as the basis for a hull-based visualization.
For each set, the convex hull of its elements is deformed to create a concave shape enclosing only the points belonging to the set.
Collins et al.~\cite{BubbleSets} create similar concave hulls in their Bubble Sets visualization, but in a different way.
They construct spanning trees for the sets of points and use them as a basis for a potential field.
Isocontours of the potential field form the bubble-like shapes of the Bubble Sets visualization (Figure~\ref{fig:nyc:BubbleSets}).
Vihrovs et al.~\cite{inverse-distance} and the authors of F2-Bubbles~\cite{F2-Bubbles} improve upon the potential field defined by Collins et al.
In particular, the potential field of Vihrovs et al.\ guarantees that regions of sets with no common elements do not overlap.
Figure~\ref{fig:nyc:Vihrovs} shows a visualization of Vihrovs et al.\ where the potential field is constructed only from the sets of points, not from the spanning trees, resulting in multiple shapes per set.

\cparagraph{Line-based.} Line-based approaches are more minimal than hull-based approaches and align with the minimal ink principle (Tufte's rule~\cite{tufte2001visual}). 
LineSets~\cite{LineSets} represents each set by one thick colored curve (Figure~\ref{fig:nyc:LineSets}).
Kelp Diagrams~\cite{kelp-diagrams} uses a spanning graph and connect points in a more controlled manner than LineSets.
When points form a cluster, line-based approaches are unsatisfactory as one would like to enclose such a cluster with a shape to provide a good sense of grouping~\cite{gestalt}.
KelpFusion~\cite{KelpFusion} (Figure~\ref{fig:nyc:KelpFusion}) is a hybrid between a line-based and hull-based approach.
It is similar to Kelp Diagrams, with the main visual difference being that faces of the spanning graph may be filled.

\cparagraph{Voronoi-based.} 
Line-based visualizations are particularly useful when there is an underlying visualization (base map) that one does not want to obscure.
If one is not concerned with this, and points belong to only one category, a simple visualization can be created from a Voronoi diagram~\cite{computational-geometry-book} of the data points by merging cells of points that belong to the same category.
The authors of GMap~\cite{GMap} take this approach. They create a Voronoi diagram of an augmented set of points to construct smooth regions.
MapSets~\cite{MapSets} (Figure~\ref{fig:nyc:MapSets}) is similar to GMap, but the authors ensure that the regions representing each category are connected by constructing pairwise non-overlapping spanning trees for each set, and compute a Voronoi diagram based on those trees.

Geiger et al.~\cite{ClusterSets} recently introduced ClusterSets, which can be drawn in all three styles: line-based, hull-based and Voronoi-based.
The visualization is based on a proximity graph; Geiger et al.\ suggest the use of $\beta$-skeletons. 
This proximity graph connects only points of the same category.
Geiger et al.\ compute a planar spanning forest of the proximity graph, minimizing the number of connected components.
The planar spanning forest can be drawn directly, yielding a line-based visualization.
However, the authors suggest adding back edges of the proximity graph and drawing the boundary polygons as in Figure~\ref{fig:nyc:ClusterSets}.
They also propose a Voronoi-based drawing where the Voronoi diagram of the polygons is computed and drawn in appropriate colors.

\cparagraph{Hypergraph drawing.}
A set visualization can be viewed as a drawing of a hypergraph whose hyperedges are the sets.
Several papers study the problem of drawing a spatial hypergraph, a hypergraph whose vertices have a fixed position, as a colored spanning graph. Specifically, they investigate ink minimization of such graphs, both in the setting where crossings are allowed~\cite{multi-colored-spanning-graphs, colored-spanning-graphs} and the setting where the graph is required to be planar~\cite{short-plane-supports}. 
These results are particularly relevant to techniques like Bubble Sets, LineSets, Kelp Diagrams and KelpFusion which use such structures.

\cparagraph{Multiple vs single shape.}
The visualizations depicted in the bottom row of Figure~\ref{fig:nyc} all use one shape per category. Hence they generally satisfy \ref{c:continuation} well, although to different degrees: LineSets provide less continuation than KelpFusion~\cite{kelp-diagrams}.
However, by using a single shape, the four visualizations inevitably cover parts of the map that are far away from points, and thereby also distort point density (\ref{c:distortion}).
Moreover, the left three visualizations (Bubble Sets, LineSets, and KelpFusion) create many intersections between shapes, which clutter the visualizations, increasing the cognitive load (\ref{c:cognitive}).
MapSets create no intersections but are complex and highly irregular.
In addition, they heavily obfuscate the base map (\ref{c:obfuscation}) and distort point density (\ref{c:distortion}).
The visualizations in the top row of Figure~\ref{fig:nyc} use multiple shapes, which allows them to score well on criteria \ref{c:cognitive}, \ref{c:obfuscation} and \ref{c:distortion}.
In Section~\ref{sec:discussion} we compare these visualizations on more datasets and backed by quantitative measures.

\section{SimpleSets}\label{sec:overview}
In this section we describe the overarching design decisions that underlie the two main phases of SimpleSets: constructing a partition into point patterns and drawing such a partition.
Figure~\ref{fig:overview} illustrates our pipeline; the first step of the drawing phase dilates the patterns, this step is explicitly included in the illustration for ease of understanding.
We present the algorithmic details for the two phases in Sections~\ref{sec:constructing-partition} and~\ref{sec:drawing-partition}.

A \emph{pattern} is a connected region in the plane that contains on its boundary or in its interior a set of data points; all data points in a pattern share a category.
Patterns can be 0-, 1-, or 2-dimensional and may consist of a single point only, polylines, spanning trees, spanning graphs, or polygonal regions. 
In addition to lines, patterns may be bounded by curves.
Two patterns are \emph{disjoint} if their sets of data points are disjoint.
Two patterns \emph{overlap} if their connected regions overlap.
Note that two disjoint patterns may overlap.

SimpleSets receives categorical point data as input and partitions them into disjoint patterns that do not overlap. Our drawing algorithm, however, can handle both overlapping and non-overlapping disjoint patterns.
To illustrate this fact we applied our drawing algorithm to the disjoint patterns that underlie LineSets (overlapping polylines), Bubble Sets (overlapping spanning trees), and ClusterSets (non-overlapping polygons).
{%
\renewcommand*{\HyperDestNameFilter}[1]{#1-appendix}%
Figure~\ref{app:fig:nyc-SimpleSets} of the supplementary material shows the results.
}%

\subsection{Islands and Banks}
SimpleSets partitions data points into two types of patterns that correspond to simple shapes: islands and banks.
In the following we assume, without loss of generality, that no three data points are collinear.

An \emph{island}~\cite{optimal-islands} is the convex hull of a set of data points. Islands have previously been studied by Bautista-Santiago et al.~\cite{optimal-islands}, who describe a dynamic program for finding the largest island in a set of $n$ data points in $O(n^3)$ time.
Furthermore, Dumitrescu and Pach~\cite{monochromatic-parts} study the worst-case cardinality of island partitions for different types of point sets. 
In our setting we want to use islands that capture the underlying point density (data distribution) well. Neither the largest island nor a partition with the fewest number of islands necessary support this quality criterion. Instead, we are using islands that cover their enclosed points well: data points are somewhat regularly distributed over the island and points which are not part of the island can easily be identified as such. Below we make this intuitive description more concrete.

\begin{figure}[b]
    \hspace*{\fill}
    \begin{subfigure}[t]{0.45\columnwidth}
        \centering
        \includegraphics[page=2]{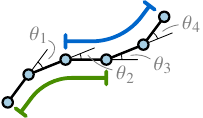}
        \caption{Turning angles $\theta_i$ and bends.}
        \label{fig:patterns-annotated:bank}
    \end{subfigure}
    \hfill
    \begin{subfigure}[t]{0.45\columnwidth}
        \centering
        \includegraphics[page=2]{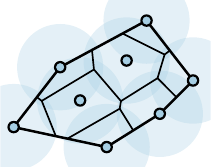}
        \caption{Cover radius.}
        \label{fig:patterns-annotated:island}
    \end{subfigure}
    \hspace*{\fill}

    \captionsetup{skip=1mm}
    \caption{Islands and banks.}
    \label{fig:patterns-annotated}
\end{figure}

\begin{figure}[tb]
    \centering
    \hspace*{\fill}
    \begin{subfigure}{0.45\columnwidth}
        \frame{\includegraphics[page=2, width=0.475\textwidth, angle=90]{figures/SimpleSets-nyc-final.pdf}}
        \vspace{-8.5mm}
        \caption{}
    \end{subfigure}
    \hfill
    \begin{subfigure}{0.45\columnwidth}
        \frame{\includegraphics[width=0.475\textwidth, angle=90]{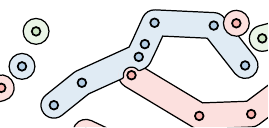}}
        \vspace{-8.5mm}
        \caption{}
    \end{subfigure}
    \hspace*{\fill}
    \captionsetup{skip=1mm}
    \caption{Banks with a limited (a) or unlimited (b) number of bends.}
    \label{fig:bendingbanks}
\end{figure}

We define a \emph{bank} as a polyline whose vertices are data points (see Figures~\ref{fig:patterns-annotated:bank} and \ref{fig:bendingbanks}). 
We characterize the complexity of a bank via its number of \emph{bends}. Specifically, each triple of consecutive points $x, y, z$  on a bank defines a signed angle (in the interval $(-\pi, \pi]$) between vectors $\overrightarrow{xy}$ and $\overrightarrow{yz}$. A positive sign indicates a counter-clockwise turn. We define a \emph{bend} in a bank as a maximal subsequence of data points such that the turning angles of all consecutive triples of data points within the sequence have the same sign. Figure~\ref{fig:bendingbanks} illustrates banks with lower or higher complexity (number of bends). 

For SimpleSets we decided to limit the number of bends of a bank to two. Furthermore, we limit the sums of turning angles of our banks to 180 degrees and the maximum turning angle to 70 degrees. We chose these values after an exploratory evaluation on our datasets; other data values might serve other datasets better.

We now attempt to formalize the intuitive notion that islands and banks cover the data points well. Recall the quality criteria that we discussed in Section~\ref{sec:related}. There is a trade-off between strong continuation of shapes depicting the same set (\ref{c:cognitive}) and the distortion of point density (\ref{c:distortion}) and the occlusion of a potential underlying visualization (\ref{c:obfuscation}). To point: a set visualization that draws only the data points results in no distortion and does not occlude a possible underlying visualization.
However, there is little continuation between shapes that depict the same set. On the other hand, every visualization that adds any form of visual continuation will inevitably incur distortion and occlusion.

In SimpleSets, we quantify this trade-off as follows.
We say that an island $I$ is \emph{$r$-covered} if its convex hull, including its interior, is covered by disks of radius $r$ centered at data points in $I$.
A bank is \emph{$r$-covered} if the edge lengths of its polyline are bounded by $2r$.
Computationally, we can verify whether an island $I$ is $r$-covered for some $r$ in $O(|I| \log |I|)$ time. 
To do so, we compute a Voronoi diagram of $I$, and take the intersection with the convex hull of $I$ (Figure~\ref{fig:patterns-annotated:island}).
The result is an arrangement consisting of Voronoi cells, some of which are clipped by the convex hull.
Junctions where edges meet are called vertices.
There are $O(|I|)$ vertices, and for each we know the distance to its closest points in $I$.
Island $I$ is $r$-covered if, and only if, all vertices have distance at most $r$ to their closest point in $I$.

The \emph{cover radius} $c(P)$ of a pattern $P$ is the smallest $r$ such that $P$ is $r$-covered.
If the cover radius limit is zero, then the only patterns are single data points.
Conversely if the cover radius is not restricted, then any island or bank forms a pattern regardless of its distribution of data points.
In between these two extremes, we find island and banks that form a spatial pattern in the sense that no location within the shape has distance (measured within the shape) larger than $r$ to a data point.

\begin{figure}[tbh]
    \centering
    \includegraphics{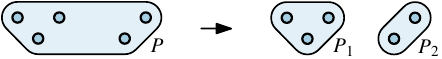}
    \caption{Island $P$ is not regular as it can be split into patterns $P_1$ and $P_2$ with $\max(c(P_1), c(P_2))$ considerably smaller than $c(P)$.}
    \label{fig:regular-islands}
\end{figure}

In Section~\ref{sec:constructing-partition} we describe our algorithm that incrementally constructs a partition into islands and banks by merging patterns with ever increasing cover radius. We prefer patterns that are \emph{regular} in the sense that all points inside the pattern are somewhat equally distributed and do not form clearly discernible sub-patterns (see Figure~\ref{fig:regular-islands}). This regularity can be quantified via the covering radius; in the next section we show how our preference can be incorporated in our incremental algorithm. 

\begin{figure}[th]
    \centering
    \includegraphics{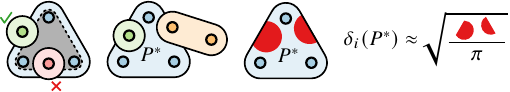}
    \captionsetup{skip=2mm}
    \caption{Left: grey shows the region where no points of other patterns may lie. Middle-left: undesirable positioning of other patterns. Middle-right: intersection area of $P^*$ with singleton points. Right: delay applied to events that would create $P^*$.}
    \label{fig:avoid}
\end{figure}

Last but not least: we want our islands and banks to be clearly separated from other data points. After partitioning the input into patterns (islands and banks) our algorithm dilates all patterns with a \emph{dilation radius} $d_r$ and then draws the dilated patterns (see Figure~\ref{fig:overview}). The patterns we create are disjoint by construction. However, we also want to avoid patterns that contain other points ``too deep'' inside their dilated shape. Naturally we cannot avoid input points that lie very close together. However, we will never create patterns that are too close to another point, which we quantify as half the dilation radius~$d_r$ (see Figure~\ref{fig:avoid} left, note that points are drawn with a circle of size $\approx d_r/3$). Patterns that contain other points within their dilated shape but at distance greater than $d_r/2$ are admissible but not desirable (see Figure~\ref{fig:avoid} middle); as with irregular islands we encourage our algorithm to not create them using a delay function. We quantify this so-called \emph{intersection delay} using the area of overlap between a potential pattern and other existing patterns (see Figure~\ref{fig:avoid} right). The technical details are described in Section~\ref{sec:constructing-partition}.

\subsection{Enclosing Shapes}\label{sec:enclosing-shapes}

The drawing algorithm of SimpleSets takes as input a set of disjoint, but possibly overlapping, patterns and computes an \emph{enclosing shape} for each of them. When choosing the enclosing shape for SimpleSets, we considered four factors. First of all, we were of course looking for a \emph{simple} shape. We consider a shape to be simple, if its boundary is smooth and consists of few geometric primitives. Second, we want the enclosing shape to be quite \emph{similar} to the patterns we find in the data. Third, we want to use an enclosing shape that clearly \emph{contains} the data points of the corresponding pattern, and fourth the enclosing shape should equally clearly \emph{exclude} any data points that are not part of the pattern. In the previous subsection we already discussed how we can facilitate containment and exclusion when choosing patterns; here we discuss further design decisions that reinforce the results when drawing the enclosing shapes.

For SimpleSets we chose the Minkowski sum of a disk of the dilation radius $r_d$ with the region of the pattern as the basis for our enclosing shape (see Figure~\ref{fig:overview}).
We refer to the resulting shapes as \emph{dilated patterns}.
Dilated patterns are simple according to our definition: their boundary is smooth and consists only of straight segments and circular arcs. Furthermore, by design dilated patterns are similar to their corresponding pattern and contain its data points in its interior (with a distance of at least $d_r$ to the boundary). However, dilated patterns may contain data points in their interior (close to their boundary) which are not part of their pattern. This might occur because data points of different categories lie at a distances less than $d_r$ or because we created an admissible but not desirable pattern (see the previous subsection). Hence, we need to find a consistent way to modify overlapping dilated patterns such that 
the set containment is clear for every data point.

\begin{figure}[b]
    \centering
    \hspace*{\fill}
    \begin{subfigure}{0.45\columnwidth}
        \centering
        \includegraphics[page=2, width=0.85\textwidth]{overlap-example-1-new.pdf}
        \caption{Dilated patterns}
        \label{fig:dilated-patterns}
    \end{subfigure}
    \hfill
    \begin{subfigure}{0.45\columnwidth}
        \centering
        \includegraphics[page=4, width=0.85\textwidth]{overlap-example-1-new.pdf}
        \caption{Voronoi}
        \label{fig:drawing-comparison:Voronoi}
    \end{subfigure}
    \hspace*{\fill}
    
    \hspace*{\fill}
    \begin{subfigure}{0.45\columnwidth}
        \centering
        \includegraphics[page=3, width=0.85\textwidth]{overlap-example-1-new.pdf}
        \caption{Geodesic hull}
        \label{fig:drawing-comparison:geodesic}
    \end{subfigure}
    \hfill
    \begin{subfigure}{0.45\columnwidth}
        \centering
        \includegraphics[page=1, width=0.85\textwidth]{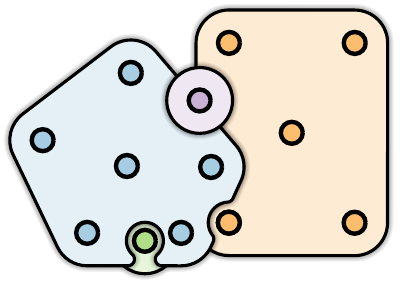}
        \caption{SimpleSets}
        \label{fig:drawing-comparison:SimpleSets}
    \end{subfigure}
    \hspace*{\fill}
    \captionsetup{skip=1mm}
    \caption{Enclosing shapes that resolve overlaps.}
    \label{fig:drawing-comparison}
\end{figure}

\cparagraph{Overlapping dilated patterns.} 
Figure~\ref{fig:drawing-comparison} shows three different methods to resolve overlapping dilated patterns. First of all, we could use a Voronoi diagram of the patterns and intersect each dilated pattern with the corresponding Voronoi cell (Figure~\ref{fig:drawing-comparison:Voronoi}). However, such a drawing may contain very little visible area around a data point and hence can make it rather difficult to determine set containment for a data point. Furthermore, the Voronoi edges are parabolic arcs which make the enclosing shapes arguably less simple and less similar to their patterns.

Another option is to make use of an implicit third dimension and \emph{stack} the enclosing shapes. Here we rely on the natural ability of the human viewer to complete contours that are partially occluded~\cite{gestalt}. A stacked drawing provides us, in a sense, with more space to work with as a data point can be clearly outside an enclosing shape even if its position in the drawing is contained in the (implied) enclosing shape. See, for example, Figure~\ref{fig:overview} right: a blue point lies inside the dilated green island pattern, but since the blue bank shape has been visually stacked above the green enclosing shape, it is still clear that this point belongs to the blue bank only.

When stacking dilated patterns we might still need to deform the boundary of an upper pattern to ensure the clear visibility of a point in a lower pattern. One option to do so are geodesic hulls, which are illustrated in Figure~\ref{fig:drawing-comparison:geodesic}. Because geodesic hulls use bitangents between data points, the enclosing shape might be deformed with respect to the dilated pattern even far away from overlaps. Since the bitangents can have any orientation, they also negatively impact the similarly of the enclosing shape to its pattern. For SimpleSets we decided to use smoothed circular cut-outs in the upper enclosing shapes to expose points in the lower shapes (see Figure~\ref{fig:drawing-comparison:SimpleSets}). Our enclosing shapes use only straight lines that are parallel to the boundary of their patterns, as well as circular arcs. This design, we believe, induces a clean look and feel and clearly shows the set containment of each data point.

\begin{figure}[tb]
    \centering
    \hspace*{\fill}
    \begin{subfigure}{0.45\columnwidth}
        \centering
        \includegraphics[page=1,width=\textwidth]{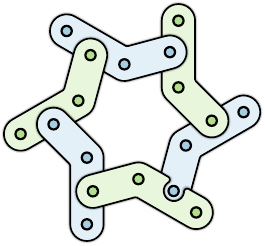}
        \caption{Stacking}
    \end{subfigure}
    \hfill
    \begin{subfigure}{0.45\columnwidth}
        \centering
        \includegraphics[page=2,width=\textwidth]{hexaswirl.pdf}
        \caption{Physically realizable}
    \end{subfigure}
    \hspace*{\fill}
    \captionsetup{skip=1mm}
    \caption{Example instance where the physically realizable model allows better enclosing shapes than the stacking model.}
    \label{fig:physically-realizable}
\end{figure}

\cparagraph{Stacking dilated patterns.}
Since we decided to stack enclosing shapes, we need to determine a drawing order in the regions where the dilated patterns overlap each other.
A simple approach would be to determine a total order on the shapes and stack them in this order. However, we choose the more powerful physically realizable model \cite{proportional-symbol-maps} that has no global stacking order, but instead only stacks shapes locally (Figure~\ref{fig:physically-realizable}).
More precisely, we allow an enclosing shape to be drawn on top of another in one area, but below that same other shape in a different area, as long as this can be achieved by ``bending''  the shapes in the implied third dimension; cutting shapes is not allowed. 
Physically realizable stacking is heavily utilized in the SimpleSets drawings of LineSets and Bubble Sets 
{%
\renewcommand*{\HyperDestNameFilter}[1]{#1-appendix}%
in Figure~\ref{app:fig:nyc-SimpleSets} of the supplementary material.
}%

\begin{figure}[b]
    \centering
    \begin{subfigure}{0.32\columnwidth}
        \centering
        \includegraphics[page=3]{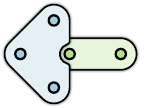}
    \end{subfigure}
    \hfill
    \begin{subfigure}{0.32\columnwidth}
        \centering
        \includegraphics[page=9]{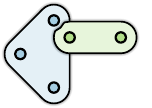}
    \end{subfigure}
    \hfill
    \begin{subfigure}{0.32\columnwidth}
        \centering
        \includegraphics[page=6]{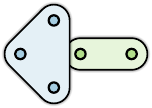}
    \end{subfigure}

    \vspace{2mm}

    \begin{subfigure}{0.32\columnwidth}
        \centering
        \includegraphics[page=2]{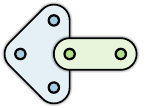}
    \end{subfigure}
    \hfill
    \begin{subfigure}{0.32\columnwidth}
        \centering
        \includegraphics[page=8]{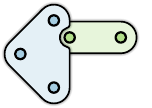}
    \end{subfigure}
    \hfill
    \begin{subfigure}{0.32\columnwidth}
        \centering
        \includegraphics[page=5]{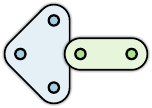}
    \end{subfigure}
    
    \caption{Stacking preferences, preferred stacking at the bottom. 
    Left: avoids modification of shape; midde: avoids modification of circular arc; right: avoids covering of circular arc.
    }
    \label{fig:pairwise-preferences}
\end{figure}

When deciding on the best local stacking orders, we follow a few simple principles, based on the assumption that humans can visually complete interrupted or modified line segments better than circular arcs (see Figure~\ref{fig:pairwise-preferences}). First and foremost, we prefer orders which do not require us to introduce cut-outs to avoid points. If we cannot avoid adding a cut-out, then we prefer to modify a line segment over modifying a circular arc. Finally, if we need to cover a piece of an enclosing shape, then we prefer to cover a line segment over covering a circular arc. In Section~\ref{sec:drawing-partition} we explain how we capture these preferences in our drawing algorithm.

\section{Partitioning Algorithm}\label{sec:constructing-partition}

SimpleSets uses a greedy incremental clustering algorithm to partition categorical point data into disjoint and non-overlapping islands and banks. 
We initialize our algorithm with a partition where each data point forms a separate pattern; afterwards we iteratively merge patterns. Whenever our algorithm merges two source patterns $P_1$ and $P_2$ into a target pattern $P^*$, we ensure that the data points and regions of $P_1$ and $P_2$ are subsets of the data points and region of $P^*$. As a result, the size of the regions covered by patterns monotonically increases as our algorithm proceeds.
We store all intermediate partitions; the sequence of partitions forms a topological filtration and can also be interpreted as a hierarchical agglomerative clustering on the input points.

Our algorithm performs a discrete event simulation, starting at time $t=0$ with the initial disjoint patterns. As $t$ increases, we merge patterns in the partition into larger patterns. We maintain the following invariant:

\begin{itemize}[noitemsep]
    \item The patterns in the partition are disjoint and do not overlap.
    \item All patterns are $t$-covered.
\end{itemize}
Thus the time parameter $t$ is a single intuitive parameter that allows the user to select the scale at which they wish to define patterns. 
{%
\renewcommand*{\HyperDestNameFilter}[1]{#1-appendix}%
In Figure~\ref{app:fig:times} of the supplementary material
}%
we illustrate the influence of $t$ via four SimpleSets visualizations of one dataset that are constructed from partitions at different times in the sequence.

There is only one type of event in our simulation: the \emph{merge event}. A merge event is associated with two source patterns $P_1$ and $P_2$ to be merged, the target pattern $P^*$ that is the result of the merge, and the time $t$ at which the merge would take place. We maintain the events in a priority queue on the event times and handle them in order of priority; two events that occur at the same time are handled in arbitrary order.

\begin{figure}[b]
    \hspace*{\fill}
    \begin{subfigure}{0.45\columnwidth}
        \centering
        \includegraphics[page=1]{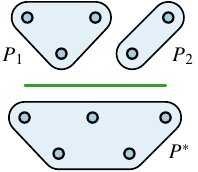}
        \caption{Good merge}
        \label{fig:cover-radius-delay:good}
    \end{subfigure}
    \hfill
    \begin{subfigure}{0.45\columnwidth}
        \centering
        \includegraphics[page=2]{cover-radius-delay.pdf}
        \caption{Mediocre merge}
        \label{fig:cover-radius-delay:bad}
    \end{subfigure}
    \hspace*{\fill}
    \captionsetup{skip=1mm}
    \caption{The target patterns $P^*$ in {\upshape (a)} and {\upshape(b)} have identical cover radius $c(P^*)$. But the pattern in {\upshape(b)} is less regular and has more pronounced sub-patterns. Our discrete event simulation delays merge {\upshape(b)}.}
    \label{fig:cover-radius-delay}
\end{figure}

Below we describe in detail how we create and handle events in our discrete event simulation. In principle, the time $t$ of a specific event corresponds directly to the covering radius that allows the corresponding pattern to be formed. However, as discussed in Section~\ref{sec:overview}, we prefer regular islands, which contain points that are somewhat equally distributed, over patterns that exhibit clear sub-patterns (see Figure~\ref{fig:cover-radius-delay}). Hence we add a processing delay to those events that would create islands which are not regular. This delay increases the time at which the event would occur and thereby increases the chance that the corresponding pattern is never formed, since other merge events might already have used a source pattern, or have blocked the target pattern, by the time the delayed event occurs. 

We also prefer patterns that are well-separated from other points. Patterns that are too close to another point (distance less than half the dilation radius $r_d$) are never created by our algorithm, since the dilated pattern would enclose that unrelated point too much. Patterns that keep a distance of less than $r_d$ to other points result in dilated patterns that still enclose those unrelated points, but less deeply. As discussed in Section~\ref{sec:overview}, these patterns are admissible but not desirable; we hence add a delay to the events that would create them. 

\cparagraph{Creating events.}
Our algorithm creates one or multiple events for two source patterns $P_1$ and $P_2$ as follows.
If either $P_1$ or $P_2$ is an island, then we create one event for them to merge into an island $P^*$ which is the convex hull of the data points contained in $P_1$ and $P_2$.
To maintain the invariant of our algorithm and ensure that the sequence of patterns is indeed a filtration, we do not allow islands to merge into a bank.

If both $P_1$ and $P_2$ are banks, then we create up to four events that correspond to the four ways in which we can connect the two ends each of $P_1$ and $P_2$. If the polyline corresponding to a specific connection is a bank according to our definition, then it becomes a target pattern $P^*$.

As discussed above, a priori the time $t$ associated with an event is the cover radius $c(P^*)$ of $P^*$. However, a less desirable merge will incur additional delays to reduce the chance that it is actually executed.
If $P_1$ and $P_2$ both have a small cover radius then we delay merging them into a pattern $P^*$ with a large cover radius (Figure~\ref{fig:cover-radius-delay}), since that merge would create a pattern which is not regular.
Such a merge is delayed by a \emph{regularity delay} of $\delta_r(P_1, P_2, P^*) = c(P^*) - \max(c(P_1), c(P_2))$. If both $P_1$ and $P_2$ are single data points then we add no delay.
Next we measure the overlap (the area of intersection) of the target pattern $P^*$ with all dilated data points.
Let $D_U$ denote the union of dilated data points and 
let $D_1$, $D_2$ and $D^*$ denote the dilated versions of $P_1$, $P_2$ and $P^*$.
We first compute the set $A = (D^* \setminus (D_1 \cup D_2)) \cap D_U$ of overlap between $D^*$ and the parts of dilated points that are neither in $D_1$ nor in $D_2$. Then we use the area $a$ of this overlap $A$ to set the \emph{intersection delay} $\delta_i(P^*) = \sqrt{a / \pi}$ (see Figure~\ref{fig:avoid} right).

\cparagraph{Initialization.}
We initialize the algorithm with a partition where each data point is a pattern.
During initialization, we create events for every pair of distinct data points of the same category.

\cparagraph{Handling events.}
When handling an event, we first need to check if the corresponding merge is possible given the current partition into patterns. If either of the source patterns $P_1$ or $P_2$ are not part of the current partition, then the merge is not possible and we simply skip the event. Similarly, if the target pattern $P^*$ overlaps any pattern in the current partition, other than $P_1$ and $P_2$, the merge is not possible and the event is skipped.
If the merge is possible, then we remove patterns $P_1$ and $P_2$ from the current partition and replace them by pattern $P^*$. Then we create events for the new pattern $P^*$ and each of the other patterns of the same category in the current partition. If the current event was processed with a regularity or an intersection delay, then the newly created events might lie in the past. In this case they are immediately processed by our algorithm.

\cparagraph{Running time.}
We now analyse the running time of our partitioning algorithm. 
Let $n$ denote the number of data points.
We can compute the regularity and intersection delays in $O(n^2)$ time and, thus, create a new event in $O(n^2)$ time.
Furthermore, we can decide in $O(n^2)$ time whether an event must be skipped.
We initially create $O(n^2)$ events in $O(n^4)$ time. 
Our discrete event simulation modifies the partition at most $n-1$ times, since each merge reduces the number of patterns in the partition by one.
Therefore, at most $n-1$ events result in actual merges, creating $O(n)$ new events each time.
Thus, we handle in total $O(n^2)$ events, each of which takes $O(n^2)$ time.
Insertions and deletions in the priority queue take $O(\log n)$ time each.
Hence, the total running time of the partitioning algorithm is $O(n^4)$.

A running time of $O(n^4)$ is comparatively high and sounds forbidding in practice, however, set visualizations are rarely applied to sets with thousands of data points at predefined locations.
Our implementation created the SimpleSets visualization in Figure~\ref{fig:nyc} in 232 milliseconds on a laptop with an Intel(R) Core(TM) i7-11800H 2.30 GHz processor. The intersection delay has a large impact on the running time but little impact on the partitions in practice; omitting it speeds up the computation to 58 milliseconds. In Figure~\ref{fig:running-time} we plot the running times for the other examples discussed in this paper: computing SimpleSets for datasets with 500 data points still takes only about 1 second.

\cparagraph{Choosing a partition.}
As mentioned above, our partitioning algorithm returns a sequence of partitions and the user can choose a suitable partition via the time parameter $t$.
In our experiments, we found that partitions in time range $[2r_d, 6r_d]$, where $r_d$ denotes the dilation radius, capture patterns well and provide a good trade-off between connecting points and covering space.
{%
\renewcommand*{\HyperDestNameFilter}[1]{#1-appendix}%
Figure~\ref{app:fig:times} of the supplementary material
}%
uses partitions at times $2.5r_d$, $3.5r_d$, $4.5r_d$, and $6r_d$.
 
\section{Drawing Algorithm}\label{sec:drawing-partition}

\begin{figure*}[tb]
    \begin{subfigure}[t]{0.15\textwidth}
        \centering
        \includegraphics[page=1]{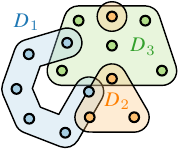}
        \caption{Dilated patterns}
        \label{fig:drawing-algorithm:dilated}
    \end{subfigure}
    \hfill
    \begin{subfigure}[t]{0.15\textwidth}
        \centering
        \includegraphics[page=4]{figures/drawing-algorithm.pdf}
        \caption{Component where $D_2$ is intersected}
    \end{subfigure}
    \hfill
    \begin{subfigure}[t]{0.16\textwidth}
        \centering
        \includegraphics[page=3]{figures/drawing-algorithm.pdf}
        \caption{Components where $D_1$ and $D_3$ intersect}
    \end{subfigure}
    \hfill
    \begin{subfigure}[t]{0.225\textwidth}
        \centering
        \includegraphics[page=5]{figures/drawing-algorithm.pdf}
        \caption{The stacking preferences of three components meet in one face}
        \label{fig:drawing-algorithm:meet}
    \end{subfigure}
    \hfill
    \begin{subfigure}[t]{0.1\textwidth}
        \centering
        \includegraphics[page=6]{figures/drawing-algorithm.pdf}
        \caption{Graph of (d) and total order}
        \label{fig:drawing-algorithm:graph}
    \end{subfigure}
    \hfill
    \begin{subfigure}[t]{0.06\textwidth}
        \centering
        \includegraphics[page=7]{figures/drawing-algorithm.pdf}
        \caption{Modify part of $D_2$}
    \end{subfigure}
    \captionsetup{skip=1mm}
    \caption{An overview of the drawing algorithm. See Figure~\ref{fig:overview} for the input patterns and the final rendering. In this example, the hypergraph $G$ consists of only one non-singleton hyperedge, shown in (d), as there is only one face in the arrangement that is contained in at least three dilated patterns.}
    \label{fig:drawing-algorithm}
\end{figure*}

The output of the partitioning algorithm described in Section~\ref{sec:constructing-partition} is a set of disjoint patterns that do not overlap. Our drawing algorithm can also handle sets of disjoint, but possibly overlapping patterns. In the supplementary material we illustrate this by rendering patterns from the state-of-the-art with our drawing algorithm. 
Disjoint but possibly overlapping patterns are, therefore, the input that we assume for this section. Figure~\ref{fig:drawing-algorithm} and Algorithm~\ref{alg:drawing} summarize our drawing algorithm.

In a first step, we dilate each pattern using the Minkowski sum with a disk of radius $d_r$ (see Figure~\ref{fig:overview}). We denote the dilated patterns by $D_1, \dots, D_m$. Even if the patterns do not overlap, the dilated patterns might (see Figure~\ref{fig:drawing-algorithm:dilated}). Wherever two or more dilated patterns overlap, we have to determine their drawing order. 
In Section~\ref{sec:overview} we discussed our preferences for stacking two enclosing shapes; in this section we explain how to combine such pairwise-preferences into local stacking orders for a set of enclosing shapes.

Once we have computed local stacking orders, we might have to modify the boundary curves of the upper shapes to ensure that data points in lower shapes remain visible. In Section~\ref{sec:overview} we discussed design alternatives for modified upper shapes and settled on circular cut-outs in otherwise straight-line convex hulls for SimpleSets. Computing these cut-outs is conceptually simple; however, several edge cases require additional care. We attempt to convey the major considerations in this section and refer the reader to the supplementary material for details.

\begin{algorithm}[tb]
\Input{Patterns $P_1, \dots, P_m$}
\Output{A drawing of $P_1, \dots, P_m$ using enclosing shapes}
\BlankLine
$D_1, \dots, D_m \gets $ dilate patterns $P_1, \dots, P_m$\;
$\mathcal{A} \gets$ arrangement of $D_1, \dots, D_m$\;
\BlankLine
\For{$1 \leq i < j \leq m$ } { 
    \For {$C \in \mathcal{C}_{i, j}$} {
        create a relation $R_{i, j, C} \in \{<, =, >\}$ and store a reference to $R_{i, j, C}$ in each face $f \in C$ in set $R(f)$\;
        compute the stacking preference of $i$ and $j$ in $C$ and set $R_{i, j, C}$ accordingly\;
    }
}
\BlankLine
$G(V, E) \gets $ hypergraph where relations $R_{i, j, C}$ form $V$ and $E$ consists of maximal sets of relations present in a face\;
\BlankLine
\For{$e \in E$} {
    determine a stacking order for $e$ and set each $R_{i, j, C} \in e$ to $<$ or $>$ appropriately\;
}
\BlankLine
\For{face $f$ in $\mathcal{A}$} {
    determine a stacking order in $f$ based on $R(f)$\;
}
\BlankLine
\For{$1 \leq i \leq m$}{ 
    \For{$C \in \mathcal{C}_i$} {
        modify parts of $D_i$ in $C$ such that data points of shapes below $i$ in the order stored in $f \in C$ are exposed\;
    }
}
\BlankLine
draw each face of $\mathcal{A}$
\caption{Drawing enclosing shapes}
\label{alg:drawing}
\end{algorithm}

\cparagraph{Computing stacking orders.} We formally approach this problem as follows, see Figure~\ref{fig:drawing-algorithm} for illustrations. 
As a first step, we build the arrangement~$\mathcal{A}$~\cite{computational-geometry-book} induced by the dilated patterns.
Every face $f$ of $\mathcal{A}$ stores a set $D(f)$ of dilated patterns that contain the face.
Let $\mathcal{C}_{i, j}$ denote the set of connected components of faces $f$ in $\mathcal{A}$ such that $D_i$ and $D_j$ are part of $D(f)$. 
Let $\mathcal{C}_i$ be the set of connected components of faces $f$ in $\mathcal{A}$ where $D_i \in D(f)$ and $|D(f)| \geq 2$.
For each pair of dilated patterns $D_i, D_j$ and component $C \in \mathcal{C}_{i, j}$ we create a relation $R_{i, j, C}$ that takes a value in $\{<, =, >\}$, which indicates the order in which $D_i$ and $D_j$ are stacked in component $C$.

Initially $R_{i, j, C}$ contains the stacking preference according to the simple rules described in Section~\ref{sec:overview}.
If no rule applies then there is no preference, indicated by $=$.
After this initial step, every face $f$ in the arrangement contains a set of relations $R(f)$.
For our drawing, we need for every face $f$ a stacking order on $D(f)$ such that the orders of all faces together form a physically realizable drawing. 
To achieve this, we set each relation $R_{i, j, C}$ to $<$ or $>$, and derive a stacking order in $f$ from $R(f)$.
Faces where three or more dilated patterns coincide contain more than one relation and impose constraints because the relations in a face should not form a cyclic dependency (Figure~\ref{fig:drawing-algorithm:meet}).

We capture these constraints in a hypergraph on the set of all relations $R_{i, j, C}$.
Each hyperedge will be a set of relations that meet in a face and should not form a cyclic dependency.
If $R(f) \subseteq R(f')$ for distinct faces $f$ and $f'$, then $f$ can inherit the ordering computed for $f'$.
Hence, the edges of the hypergraph are
$E = \{ R(f) \mid R(f) \text{ is maximal} \}.$
Consider a hyperedge $e \in E$.
We compute a total ordering of the dilated patterns that are part of the relations of this hyperedge by constructing a graph $G_e$.
The vertices of~$G_e$ are the set of all indices of dilated patterns that are present in the relations stored in $e$.
An edge is present in $G_e$, directed from $i$ to $j$, if a relation indicates that $D_i$ should be stacked on top of $D_j$.
If $G_e$ has no cycles then we use a topological order of $G_e$ and set the relations $R_{i, j, C}$ to match this total order (Figure~\ref{fig:drawing-algorithm:graph}).

If $G_e$ has cycles, then we remove them by flipping edges. We want to minimize the number of flipped edges, since they correspond to drawing orders that go against our rules. The corresponding computational problem is challenging, but brute-force solutions are computationally feasible for  standard set visualizations. However, in all our experiments with SimpleSets we have not encountered a single example where $G_e$ has a cycle. Stronger yet, in all our SimpleSets examples there is a total stacking order of enclosing shapes which satisfies our rules.

\begin{figure*}[tb]
    \begin{subfigure}[t]{0.13\textwidth}
        \centering
        \includegraphics[scale=1, page=1]{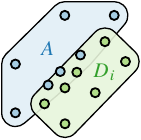}
        \caption{Overlap}
        \label{fig:curve-modification:start}
    \end{subfigure}
    \hfill
    \begin{subfigure}[t]{0.13\textwidth}
        \centering
        \includegraphics[scale=1.5, page=4]{intertwined-morph-fresh.pdf}
        \caption{Disks $X$ and $Y$}
        \label{fig:curve-modification:disks}
    \end{subfigure}
    \hfill
    \begin{subfigure}[t]{0.16\textwidth}
        \centering
        \includegraphics[scale=1.5, page=6]{intertwined-morph-fresh.pdf}
        \caption{Disks $X$ expanded by $r_s$}
        \label{fig:curve-modification:expanded}
    \end{subfigure}
    \hfill
    \begin{subfigure}[t]{0.17\textwidth}
        \centering
        \includegraphics[scale=1.5, page=9]{intertwined-morph-fresh.pdf}
        \caption{Convex hulls~\cite{DBLP:journals/comgeo/Rappaport91} of disk components in (c) minus $Y$}
        \label{fig:curve-modification:components}
    \end{subfigure}
    \hfill
    \begin{subfigure}[t]{0.11\textwidth}
        \centering
        \includegraphics[scale=1.5, page=10]{intertwined-morph-fresh.pdf}
        \caption{Component $C$ after subtraction}
        \label{fig:curve-modification:modified}
    \end{subfigure}
    \hfill
    \begin{subfigure}[t]{0.13\textwidth}
        \centering
        \includegraphics[scale=1, page=2]{intertwined-morph-fresh.pdf}
        \caption{Final drawing after smoothing}
        \label{fig:curve-modification:final}
    \end{subfigure}
    \captionsetup{skip=1mm}
    \caption{Modification of dilated patterns to expose points beneath.
    Figures (b)--(e) show a closeup of the component $C$ in (a) and (f). 
    }
    \label{fig:curve-modification}
\end{figure*}

\cparagraph{Curve modification.}
Once the local stacking orders have been determined, we modify the dilated patterns where needed to expose data points inside lower shapes.
This process is summarized in Figure~\ref{fig:curve-modification}.
Consider an arbitrary component $C \in \mathcal{C}_i$ for some $i$.
Let $A$ be the set of patterns that are below $D_i$ in the stacking order in the faces of $C$. 
We expose data points of patterns in $A$ by modifying $D_i$.

Our algorithm aims to keep a disk of radius $r_c = {5}/{8} \cdot r_d$ centered at each element visible.
To this end, we place exclusion (red) disks of radius zero at each data point of each pattern in $A$.
We also place inclusion (green) disks of radius zero at each data point of~$P_i$.
Then we grow the disks at a uniform rate.
If two disks of different color collide then we stop their growth.
We also stop the growth of a red disk when it reaches $r_c$ and the growth of a green disk when it reaches $r_d$.

The result of this growth process is a set $X$ of exclusion disks and a set $Y$ of inclusion disks such that any disk in $X$ is disjoint from those in $Y$ and vice versa (Figure~\ref{fig:curve-modification:disks}).
We now cut the disks in $X$ out of $D_i$ to expose the corresponding points and smooth the result using the Minkowski sum and difference with a disk of radius $r_s = r_d / 5$.
When disks in $X$ are sufficiently close (Figure~\ref{fig:curve-modification:expanded}) then we cut out their convex hull~\cite{DBLP:journals/comgeo/Rappaport91} instead.
We ensure that the disks in $Y$ lie in the final pattern by not cutting them out (Figure~\ref{fig:curve-modification:components}).

\cparagraph{Running time.} The arrangement has complexity $O(n^2)$ and 
can be computed in $O(n^2 \log n)$ time using a sweep-line algorithm.
The other parts of the algorithm run in $O(n^2)$ time.
The partitioning algorithm clearly dominates the running time.

\section{Evaluation}\label{sec:discussion}
We evaluate SimpleSets based on the four criteria of Dinkla et al.~\cite{kelp-diagrams} stated in Section~\ref{sec:related}. Below, we introduce measures to capture these criteria, allowing for a quantitative analysis.
We compare SimpleSets to VPF*14~\cite{inverse-distance} and ClusterSets~\cite{ClusterSets}, as these techniques also use multiple shapes to represent each set.
We create and measure two versions of VPF*14: VPF*14+ where each vertex has a large radius of influence on the potential field, and VPF*14- where each vertex has a medium radius of influence.
VPF*14+ best matches the visualizations in the paper that introduced the method; VPF*14- uses a 25\% smaller radius.

\begin{figure}[b]
\begin{tikzpicture}
\begin{axis}[
ymin = 0,
width=8cm,
height=2.75cm,
xlabel={number of points},
ylabel={ms},
only marks
]
\addplot[color=CBblue, mark=o] coordinates {
(32, 37)
(55, 47)
(89, 304)
(96, 232)
(516, 1015)
};

\addplot[color=CBorange, mark=x] coordinates {
(32, 7)
(55, 15)
(89, 69)
(96, 58)
(516, 811)
};
\end{axis} 
\end{tikzpicture}
\captionsetup{skip=1mm}
\caption{Total running time of SimpleSets in milliseconds on our datasets. Crosses show the running time without intersection delay.}
\label{fig:running-time}
\end{figure}
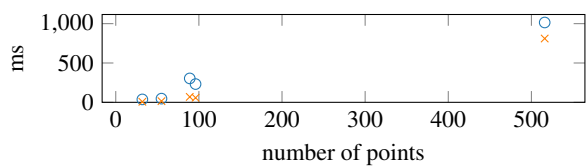

\cparagraph{Implementation.}
A prototype implementation of SimpleSets written in Kotlin is available at \href{https://github.com/tue-alga/SimpleSets}{github.com/tue-alga/SimpleSets}.
For easy access, the repository links to a web-based implementation created by compiling the Kotlin code to JavaScript.
The running time of the JVM implementation is shown in Figure~\ref{fig:running-time}.
All SimpleSets visualizations in this paper were generated by this implementation in seconds.
The drop-shadows present in some of the figures were added manually.

\cparagraph{Data.} We use the following three datasets in our evaluation. These datasets are available in the linked GitHub repository in the folder \texttt{src/commonMain/resources/example-input}.

\begin{figure*}[tb]
    \centering
    \begin{subfigure}{\columnwidth}
        \centering
        \includegraphics[width=\textwidth, page=2]{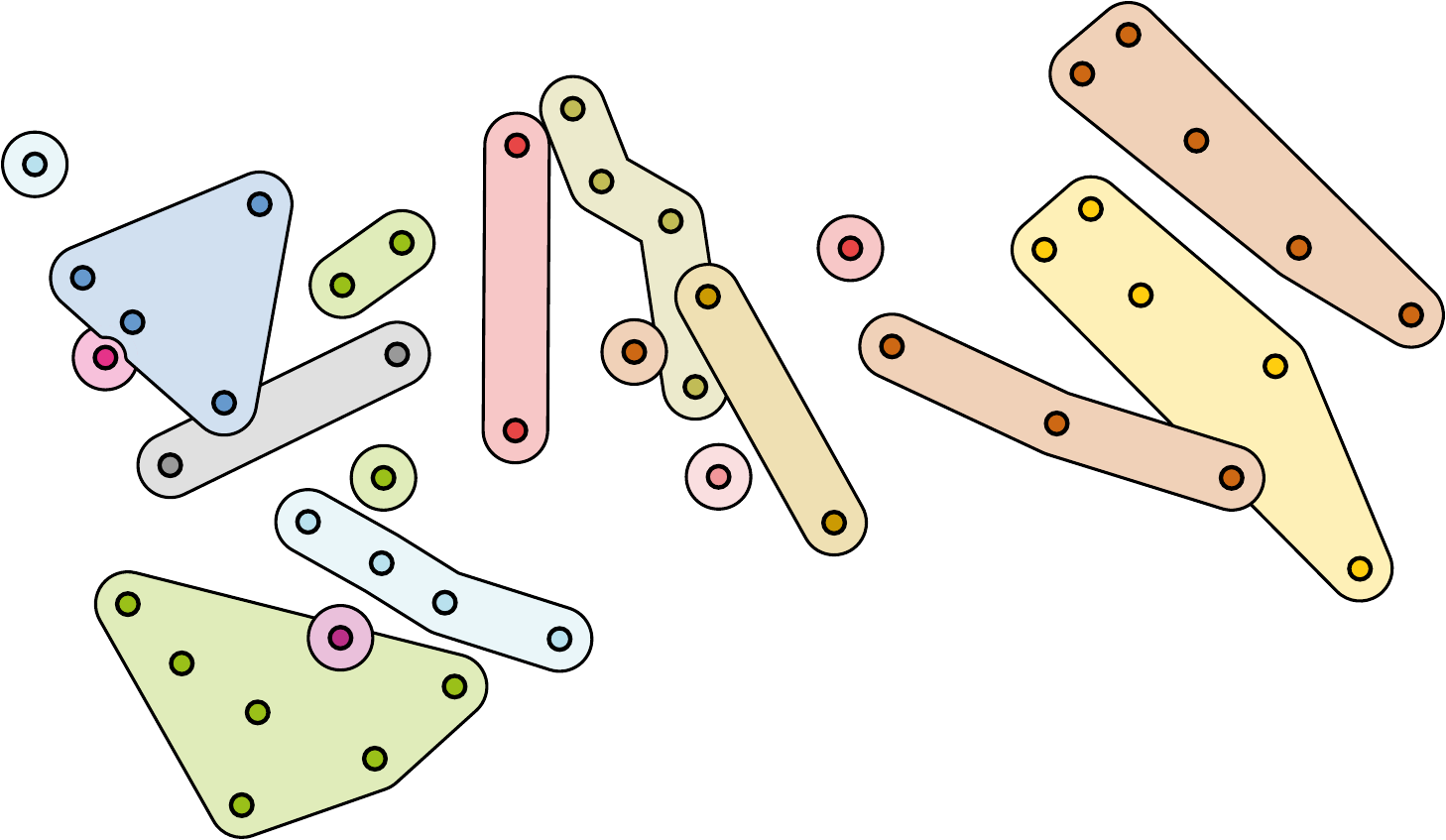}
        \vspace*{-0.75cm}
        \caption{SimpleSets}
        \label{fig:diseasome:SimpleSets}
    \end{subfigure}
    \hfill
    \begin{subfigure}{\columnwidth}
        \centering
        \includegraphics[width=\textwidth, page=2]{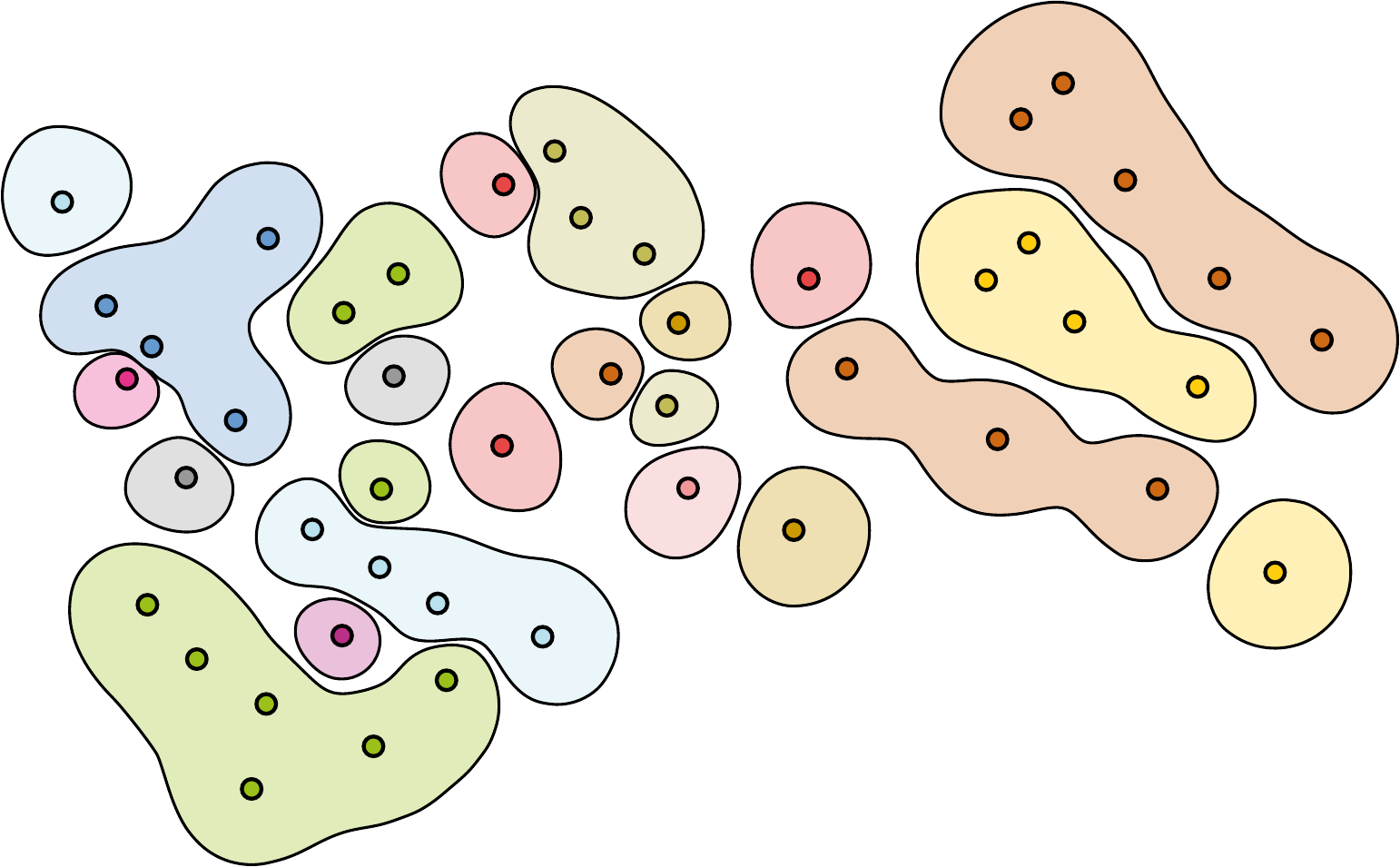}
        \vspace*{-0.75cm}
        \caption{VPF*14+~\cite{inverse-distance}}
        \label{fig:diseasome:Vihrovs}
    \end{subfigure}
    \captionsetup{skip=1mm}
    \caption{
        Part of the human disease network (HDN) dataset by Goh et al.~\cite{human-disease-network}
    }
    \label{fig:diseasome}
\end{figure*}

\begin{description}
    \item[Mills] Mills in the vicinity of Leeuwarden, The Netherlands (Figure~\ref{fig:teaser}). The dataset consists of 55 mills of four types.
    \item[NYC] Points of interest located in lower Manhattan (Figure~\ref{fig:nyc}). The dataset consists of 96 points of three types.
    \item[HDN] Vertices of an embedded graph of disorders, from the human disease network constructed by Goh et al.~\cite{human-disease-network}. Figure~\ref{fig:diseasome} shows an extract. The full dataset consists of 516 vertices (disorders) of twenty-one disorder classes. We follow Vihrovs et al.~\cite{inverse-distance} in using the graph layout of a poster by Bastian and Heymann archived at \url{https://web.archive.org/web/20121116145141/http://diseasome.eu/data/diseasome_poster.pdf}.
\end{description}

\cparagraph{Measures.}
We measure criterion~\ref{c:cognitive}, cognitive load, by measuring various aspects of shape complexity.
The complexity of shapes is difficult to measure.
A wide range of measures have been proposed and this is still an active research topic~\cite{attneave1957physical, DBLP:conf/gis/BrinkhoffKSB95, DBLP:journals/vc/Dai0ZMLY22, DBLP:conf/icip/PageKSRA03}.
We require measures that work for both the smooth shapes that are present in the VPF*14 visualization and the more polygonal, but still smooth, shapes of ClusterSets and SimpleSets.
We use the following four measures; we also provide citations to relevant literature.
\begin{description}
\item[Inflections~\cite{attneave1957physical, kelp-diagrams}]
For each shape, determine the number of inflection points. An inflection point is where the curvature of a shape changes sign. We measure the total number of inflection points present in a visualization.
    
\item[Perimeter ratio~\cite{DBLP:conf/gis/BrinkhoffKSB95}]
The ratio of the perimeter of a shape to the perimeter of the convex hull of the shape. We measure the average and maximum ratios among all shapes.

\item[Area ratio~\cite{DBLP:conf/gis/BrinkhoffKSB95}] 
The ratio of the area of a shape to the area of the convex hull of the shape. We measure the average and maximum ratios among all shapes.

\item[Curvature~\cite{matsumoto2019quantification}]
The total absolute curvature of the shape in radians. As all visualizations use closed shapes this is at least $2\pi$ for each shape; therefore, for better comparison we subtract $2\pi$. We measure the average and maximum value among all shapes.
\end{description}
To measure the number of inflection points and the total absolute curvature we sample equidistant points on the contours of the shapes
{%
\renewcommand*{\HyperDestNameFilter}[1]{#1-appendix}%
(Figure~\ref{app:fig:sampling}
}%
of supplementary material).
Holes of shapes also contribute to the number of inflections, the perimeter, and curvature of the shape (and negatively to its area).
The number of shapes (the measure we use for \ref{c:continuation}), also readily affects cognitive load, and should be taken into account as convex shapes score perfectly on the above measures.

We measure \ref{c:continuation}, strong continuation, by the number of shapes. This factor is evidently not the only factor affecting the visual continuation of sets; however, this is an attribute that is straightforward to measure and clearly correlates with continuation for the three visualizations we are comparing. 

The third criterion \ref{c:obfuscation} is straightforward to measure: we measure the area covered by the visualization. To normalize, we divide by the area of the bounding box of the input points and we present the result as a percentage value. Note that the value could exceed 100\% as shapes may lie partially outside the bounding box of the input points.

\begin{table*}[tb]
  \caption{%
	Quantitative comparison of SimpleSets, Vihrovs et al., and ClusterSets on three datasets.
    Where applicable, measures have the format \emph{avg / max}, showing both the average and maximum among all colors or shapes as appropriate. Lower values are better.%
  }
  \label{tab:measures}
  \scriptsize%
  \centering%
  \begin{tabular}{%
  	  r%
        l%
      cccccccc
  	}
  	\toprule
    & & \multicolumn{4}{c}{\ref{c:cognitive}} & \ref{c:cognitive}, \ref{c:continuation} & \ref{c:obfuscation} & \multicolumn{2}{c}{\ref{c:distortion}}\\
    \cmidrule(lr){3-6}
    \cmidrule(lr){7-7}
    \cmidrule(lr){8-8}
    \cmidrule(lr){9-10}
    & & Inflections & Perimeter ratio & Area ratio & Curvature & Shapes & Covered area & Density distortion & Cover radius \\
  	\midrule
    \multirow{4}{*}{\rotatebox{90}{Mills}} %
    & SimpleSets      & 8   & 1.01 / 1.25 & 1.07 / 2.04 & 0.72 / 5.86 & 26  & 19.4\% & 6.11\% / 11.7\% & 5.00 / 16.4 \\
    & VPF*14+         & 45  & 1.03 / 1.20 & 1.06 / 1.33 & 2.61 / 13.9 & 25  & 51.1\% & 1.84\% / 3.48\% & 2.65 / 13.3 \\
    & VPF*14-         & 22  & 1.01 / 1.09 & 1.02 / 1.26 & 0.87 / 7.41 & 35  & 32.7\% & 1.68\% / 3.27\% & 1.84 / 11.2 \\
    & ClusterSets     & 18  & 1.11 / 1.40 & 1.48 / 2.67 & 5.91 / 26.3 & 8   & 48.5\% & 8.05\% / 13.7\% & 23.2 / 37.2 \\[1mm]
    \multirow{4}{*}{\rotatebox{90}{NYC}} %
    & SimpleSets      & 16  & 1.01 / 1.11 & 1.07 / 1.87 & 1.03 / 7.39 & 36  & 25.3\% & 0.89\% / 2.07\% & 6.64 / 23.4  \\
    & VPF*14+         & 60  & 1.03 / 1.20 & 1.07 / 1.51 & 3.10 / 17.1 & 36  & 55.2\% & 1.17\% / 1.75\% & 4.65 / 18.5 \\
    & VPF*14-         & 60  & 1.03 / 1.16 & 1.07 / 1.41 & 2.66 / 13.3 & 43  & 39.8\% & 0.49\% / 0.74\% & 3.31 / 18.2 \\
    & ClusterSets     & 29  & 1.19 / 1.74 & 1.57 / 3.10 & 8.29 / 30.2 & 10  & 61.7\% & 6.87\% / 10.0\% & 28.8 / 68.1 \\[1mm]
    \multirow{4}{*}{\rotatebox{90}{HDN}} %
    & SimpleSets      & 36  & 1.00 / 1.07 & 1.02 / 1.78 & 0.27 / 9.38 & 229 & 21.6\% & 1.09\% / 8.37\% & 12.5 / 78.2  \\
    & VPF*14+         & 383 & 1.02 / 2.07 & 1.04 / 1.63 & 1.79 / 110  & 258 & 36.2\% & 0.54\% / 3.13\% & 4.73 / 44.4 \\
    & VPF*14-         & 406 & 1.02 / 1.89 & 1.04 / 1.94 & 1.92 / 70.5 & 294 & 26.5\% & 0.37\% / 1.92\% & 4.49 / 39.8 \\
    & ClusterSets     & 134 & 1.08 / 2.35 & 1.20 / 4.97 & 3.30 / 88.0 & 163 & 31.7\% & 1.87\% / 10.9\% & 30.2 / 234 \\
  	\bottomrule
  \end{tabular}%
\end{table*}

Lastly, criterion~\ref{c:distortion}: distortion of point position and density.
We use the following two measures.
\begin{description}
    \item[Density distortion] The absolute difference, per set, between the fraction of input points that belong to this set, and the fraction of area covered by shapes of this set compared to the total area covered by the visualization. We measure the average and maximum value.
    \item[Cover radius] The cover radius of a pattern captures parts of a pattern that are far away from the input points it represents. We measure the average and maximum cover radius, over all patterns. Note that the cover radius, opposed to all other values we measure, is scale-dependent and, therefore, makes sense only in comparison with visualizations on the same dataset.
\end{description}

\cparagraph{Results and discussion.}
For the visualizations, refer to Figure~\ref{fig:nyc} (NYC) and supplementary material 
{%
\renewcommand*{\HyperDestNameFilter}[1]{#1-appendix}%
Figure~\ref{app:fig:mills} (Mills), 
Figure~\ref{app:fig:nyc:VPF*14-} (NYC, VPF*14-) and 
Figures~\ref{app:fig:diseasome:SimpleSets}--\ref{app:fig:diseasome:ClusterSets} (HDN).
}%
Table~\ref{tab:measures} shows the measurements of these visualizations;
for all measures, lower values indicate better performance.

SimpleSets scores better than ClusterSets on all measures except the number of shapes. 
The individual shapes of SimpleSets are simpler.
As SimpleSets uses more shapes, though, it is unclear whether the cognitive load is lower when viewing a SimpleSets visualization compared to a ClusterSets visualization.

Comparing SimpleSets to VPF*14, we see that the shapes of VPF*14 are irregular and cover considerably more area.
VPF*14+ uses roughly the same number of shapes as SimpleSets; VPF*14- uses more shapes but covers less area than VPF*14+.
Both versions of VPF*14 generally score better on our quantitative measures of \ref{c:distortion}.
The usage of a potential field by VPF*14 explains these differences. 
In VPF*14 each point has an area of influence and shapes merge only when their areas of influence touch.
In contrast, in SimpleSets two singleton patterns may use significantly less area than a bank joining the two points.
See for example the two red points in Figure~\ref{fig:diseasome} that are part of one large vertical bank in SimpleSets, but are separate shapes in VPF*14.
Though not captured in the quantitative measures, VPF*14 does distort point position and density (\ref{c:distortion}) in ways that SimpleSets does not.
In particular, the shapes used by VPF*14 depend on the distribution of points around it, which means the same pattern of points may be drawn quite differently depending on its surroundings.
This is most clear in the shapes enclosing single points: they vary greatly in size, and the data point often lies away from the centroid of the shape.
In contrast, SimpleSets encloses single points with circles with the data point at the center of the circle.
The approach of Vihrovs et al.\ does have its advantages: the bottom-left green shape in Figure~\ref{fig:diseasome:Vihrovs} in their visualization nicely follows the distribution of points and avoids creating an intersection with the shape enclosing the magenta point.
However, in general it introduces unnecessary complexity and distortion.

\section{Future Work}
The SimpleSets partitioning algorithm uses two delays, which work well for our datasets.
It would be interesting to measure the effect these delays have; furthermore, one could consider additional delays such as a delay that penalizes merges that add an inflection to a bank.
In addition, SimpleSets relies on color to convey the category to which a point or shape belongs.
Therefore, they can clearly depict only a limited number of categories.
The use of additional visual attributes to encode the categories could mitigate this limitation.

When applying the SimpleSets drawing algorithm to large overlapping patterns, such as those of LineSets and Bubble Sets (see supplementary material), there are intersections where no preference rules apply, which leads to arbitrary ``weaving'' of the drawing. In this context it may be interesting to investigate the minimization of weaving~\cite{tunnels-bridges-switches}.
We note that a user study might be insightful and could settle lingering questions that our quantitative evaluation could not answer regarding the effectiveness of SimpleSets compared to other visualizations.

We designed SimpleSets to visualize points that each belong to one category.
An interesting direction for future work would be to adapt SimpleSets to visualize points belonging to multiple categories.
Though several parts readily extend, various questions on visual design and algorithms need to be answered on the way to an automated solution.

We close with a general question for set visualizations: how can we handle datasets that contain both very dense and quite sparse areas?
Such datasets are abundant in practice, which limits the practical application of current set visualizations for categorical point data.

\clearpage

\section*{Supplementary Materials}
\label{sec:supplementary_materials}
The supplementary material contains a detailed description of the process for computing cutouts and eight supplementary figures.
{%
\renewcommand*{\HyperDestNameFilter}[1]{#1-appendix}%
Figure~\ref{app:fig:times} 
shows four SimpleSets partitions for different time parameters. 
Figure~\ref{app:fig:nyc-SimpleSets} 
illustrates that any set of disjoint patterns can be drawn with the SimpleSets drawing algorithm. 
Figure~\ref{app:fig:sampling}
}%
shows the equidistant point sampling used to measure inflections and curvature for the quantitative analysis.
The remaining figures are outputs not present in the paper that were analyzed in the quantitative evaluation.

The code is available at \href{https://github.com/tue-alga/SimpleSets}{github.com/tue-alga/SimpleSets} and has been archived at \href{https://doi.org/10.5281/zenodo.12784670}{doi:10.5281/zenodo.12784670}.

\section*{Figure Credits}
Figure~\ref{fig:nyc:Vihrovs} is based on Figure 9c of the paper by Vihrovs et al.~\cite{inverse-distance}, but has been generated using our implementation of their method.
Figures~\ref{fig:nyc:ClusterSets}--\ref{fig:nyc:MapSets} are adapted from the paper by Geiger et al.~\cite{ClusterSets}.
The backdrop map in Figure~\ref{fig:nyc} is from \href{https://www.openstreetmap.org/copyright}{openstreetmap.org}.

%% file: SimpleSets-appendix-body.tex
\begin{bibunit}

\capstartfalse
\begin{figure*}[t]
\begin{minipage}{\textwidth}
\centering
\makeatletter
{\sffamily\huge\vgtc@sectionfont%
SimpleSets: Capturing Categorical Point Patterns\\
with Simple Shapes\\
\medskip
\bfseries\LARGE Supplementary Material\\}
\vspace{14pt}
{\large\sffamily\vgtc@sectionfont%
Steven van den Broek, Wouter Meulemans, and Bettina Speckmann\\}
\vspace{10pt}
\makeatother
\includegraphics{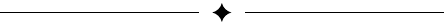}
\end{minipage}
\end{figure*}
\capstarttrue
\setcounter{figure}{0}
\setcounter{table}{0}

\section{Extended Description of the Curve Modification}\label{app:sec:drawing-partition}
\begin{figure*}[tb]
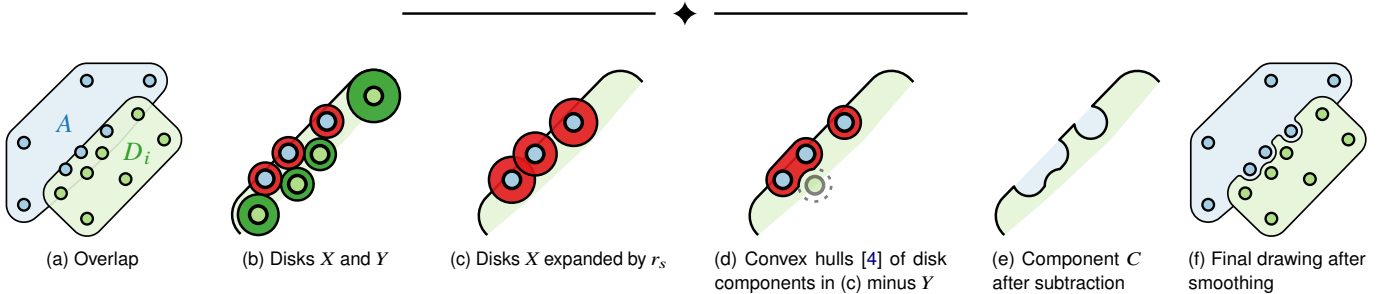

    \begin{subfigure}[t]{0.13\textwidth}
        \centering
        \includegraphics[scale=1, page=1]{intertwined-morph-fresh.pdf}
        \caption{Overlap}
        \label{app:fig:curve-modification:start}
    \end{subfigure}
    \hfill
    \begin{subfigure}[t]{0.13\textwidth}
        \centering
        \includegraphics[scale=1.5, page=4]{intertwined-morph-fresh.pdf}
        \caption{Disks $X$ and $Y$}
        \label{app:fig:curve-modification:disks}
    \end{subfigure}
    \hfill
    \begin{subfigure}[t]{0.16\textwidth}
        \centering
        \includegraphics[scale=1.5, page=6]{intertwined-morph-fresh.pdf}
        \caption{Disks $X$ expanded by $r_s$}
        \label{app:fig:curve-modification:expanded}
    \end{subfigure}
    \hfill
    \begin{subfigure}[t]{0.17\textwidth}
        \centering
        \includegraphics[scale=1.5, page=9]{intertwined-morph-fresh.pdf}
        \caption{Convex hulls~\cite{DBLP:journals/comgeo/Rappaport91} of disk components in (c) minus $Y$}
        \label{app:fig:curve-modification:components}
    \end{subfigure}
    \hfill
    \begin{subfigure}[t]{0.11\textwidth}
        \centering
        \includegraphics[scale=1.5, page=10]{intertwined-morph-fresh.pdf}
        \caption{Component $C$ after subtraction}
        \label{app:fig:curve-modification:modified}
    \end{subfigure}
    \hfill
    \begin{subfigure}[t]{0.13\textwidth}
        \centering
        \includegraphics[scale=1, page=2]{intertwined-morph-fresh.pdf}
        \caption{Final drawing after smoothing}
        \label{app:fig:curve-modification:final}
    \end{subfigure}
    \captionsetup{skip=1mm}
    \caption{Modification of dilated patterns to expose points beneath.
    Figures (b)--(e) show a closeup of the component $C$ in (a) and (f). 
    }
    \label{app:fig:curve-modification}
\end{figure*}

\cparagraph{The general approach.}
The modification process is summarized in Figure~\ref{app:fig:curve-modification}.
Consider an arbitrary component $C \in \mathcal{C}_i$ for some $i$.
Let $A$ be the set of patterns that are below $D_i$ in the stacking order in the faces of $C$. 
We expose data points of patterns in $A$ by modifying $D_i$.

Our algorithm aims to keep a disk of radius $r_c = {5}/{8} \cdot r_d$ centered at each element visible.
To this end, we place exclusion (red) disks of radius zero at each data point of each pattern in $A$.
We also place inclusion (green) disks of radius zero at each data point of~$P_i$.
Then we grow the disks at a uniform rate.
If two disks of different color collide then we stop their growth.
We also stop the growth of a red disk when it reaches $r_c$ and the growth of a green disk when it reaches $r_d$.

The result of this growth process is a set $X$ of exclusion disks and a set $Y$ of inclusion disks such that any disk in $X$ is disjoint from those in $Y$ and vice versa (Figure~\ref{app:fig:curve-modification:disks}).
We now cut the disks in $X$ out of $D_i$ to expose the corresponding points and smooth the result using the Minkowski sum and difference with a disk of radius $r_s = r_d / 5$.
When the disks in $X$ are sufficiently close (Figure~\ref{app:fig:curve-modification:expanded}) then we cut out their convex hull~\cite{DBLP:journals/comgeo/Rappaport91} instead.
We ensure that the disks in $Y$ lie in the final pattern by not cutting them out (Figure~\ref{app:fig:curve-modification:components}).

\begin{figure}[tbh]
    \hspace*{\fill}
    \begin{subfigure}{0.19\columnwidth}
        \centering
        \includegraphics[page=2, scale=0.75]{figures/3pts-bad.pdf}
        \caption{}
    \end{subfigure}
    \hfill
    \begin{subfigure}{0.3\columnwidth}
        \includegraphics[page=5, scale=0.75]{figures/3pts-bad.pdf}
        \hfill
        \includegraphics[page=1, scale=0.75]{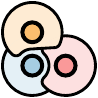}
        \caption{}
    \end{subfigure}
    \hfill
    \begin{subfigure}{0.3\columnwidth}
        \includegraphics[page=4, scale=0.75]{figures/3pts-bad.pdf}
        \hfill
        \includegraphics[page=6, scale=0.75]{figures/3pts-bad.pdf}
        \caption{}
    \end{subfigure}
    \hspace*{\fill}
    \captionsetup{skip=1mm}
    \caption{We modify the top orange pattern to expose the bottom two points by cutting the disks in $X$ out separately, even though they are close, because their intersection with the boundary of the top pattern is a circular arc. (a) Overlap. (b) Cutting out the convex hull of the two disks in $X$. (c) Cutting out the disks in $X$ separately, our preferred solution.}
    
    \label{app:fig:3pts}
\end{figure}

\begin{figure}[tbh]
    \begin{subfigure}{0.4\columnwidth}
        \centering
        \includegraphics[page=1, width=0.95\textwidth]{overlap-example-1-new.pdf}
        \caption{}
    \end{subfigure}
    \hfill
    \begin{subfigure}{0.125\columnwidth}
        \centering
        \includegraphics[page=9, width=0.9\textwidth]{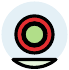}
        \caption{}
    \end{subfigure}
    \hfill
    \begin{subfigure}{0.125\columnwidth}
        \centering
        \includegraphics[page=10, width=0.9\textwidth]{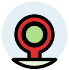}
        \caption{}
    \end{subfigure}
    \hfill
    \begin{subfigure}{0.125\columnwidth}
        \centering
        \includegraphics[page=11, width=0.9\textwidth]{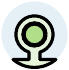}
        \caption{}
    \end{subfigure}
    \hfill
    \begin{subfigure}{0.125\columnwidth}
        \centering
        \includegraphics[page=8, width=0.9\textwidth]{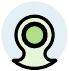}
        \caption{}
    \end{subfigure}
    \captionsetup{skip=1mm}
    \caption{(a) A setting where (b) a disk in $X$ lies fully inside $C$. (c) We cut out a thin rectangle between the circle and the closest point on the boundary of $C$. In the end, (d) the modified component is smoothed, (e).}
    \label{app:fig:edge-case:enclosed}
\end{figure}

\cparagraph{Cutting out convex hulls.}
We describe now more precisely when we group disks and cut out their convex hull.
A disk in $X$ is a candidate for a ``group cut'' only if the boundary of $D_i$ it intersects is a line segment.
Figure~\ref{app:fig:3pts} illustrates the reasoning.
A disk that does not intersect the boundary of $D_i$ at all is also a candidate, see Figure~\ref{app:fig:edge-case:enclosed} for details on this edge case.
We expand these candidate disks by increasing their radius by the smoothing radius $r_s$ (Figure~\ref{app:fig:curve-modification:expanded}).
We then determine the connected components of the intersection graph of the expanded disks. 
For each connected component, we compute the convex hull of its disks~\cite{DBLP:journals/comgeo/Rappaport91} and cut out any disk in $Y$ (Figure~\ref{app:fig:curve-modification:components}).
The result is cut out of the part of $D_i$ in component $C$ (Figure~\ref{app:fig:curve-modification:modified}).

\begin{figure}[tbh]
    \begin{subfigure}{0.49\columnwidth}
        \centering
        \includegraphics[page=12]{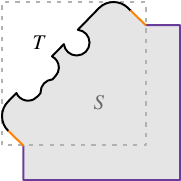}
        \caption{We extend the ends of $T$ (orange) and close it around its bounding box (purple).}
        \label{app:fig:smoothing:a}
    \end{subfigure}
    \hfill
    \begin{subfigure}{0.45\columnwidth}
        \centering
        \includegraphics[page=13]{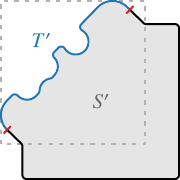}
        \caption{We cut (red) the smoothed version $S'$ of $S$ to obtain $T'$ (blue).}
        \label{app:fig:smoothing:b}
    \end{subfigure}
    \captionsetup{skip=1mm}
    \caption{Smoothing $T$ into $T'$.}
    \label{app:fig:smoothing}
\end{figure}

\cparagraph{Smoothing.}
Let $T$ be the boundary of $D_i$ in component $C$ after the cutting process; $T$ is an open curve.
We smooth by using the Minkowski sum ($\oplus$) and difference ($\ominus$) with a disk $D$ of radius $r_s = r_d/S$.
The shape of $T$ beyond the cutouts should not be smoothed. 
To ensure that, we create a closed shape $S$ by extending the endpoints of $T$ in the direction of the respective tangents and close it around its bounding box (Figure~\ref{app:fig:smoothing:a}).
The smoothed version $S'$ of $S$ is
\[
S' = (((S \ominus D) \oplus D) \oplus D) \ominus D.
\]
In terms of mathematical morphology operators~\cite{DBLP:journals/pami/HaralickSZ87}, we first apply the opening operator to smooth convex vertices and then the closing operator to smooth reflex vertices.
At the end, we cut $S'$ to obtain the smoothed boundary $T'$ (Figure~\ref{app:fig:smoothing:b}).

\section{Additional Figures}\label{app:sec:additional-figures}
\begin{figure*}[b]
    \begin{subfigure}{\columnwidth}
        \fbox{\includegraphics[width=0.97\columnwidth]{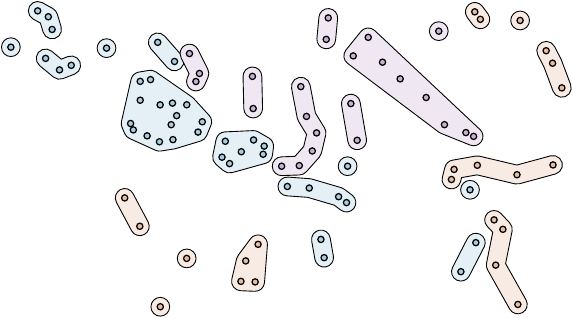}}
        \caption{Time $t = 2.5r_d$}
    \end{subfigure}
    \hfill
    \begin{subfigure}{\columnwidth}
        \fbox{\includegraphics[width=0.97\columnwidth]{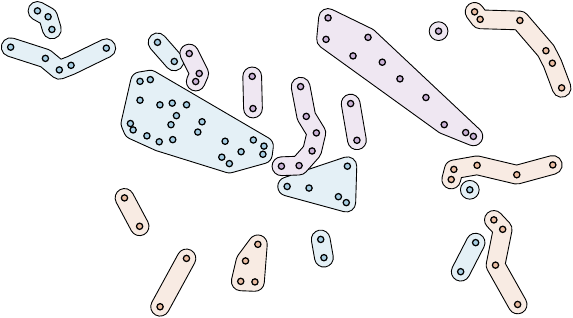}}
        \caption{Time $t = 3.5r_d$}
    \end{subfigure}
    
    \begin{subfigure}{\columnwidth}
        \fbox{\includegraphics[width=0.97\columnwidth]{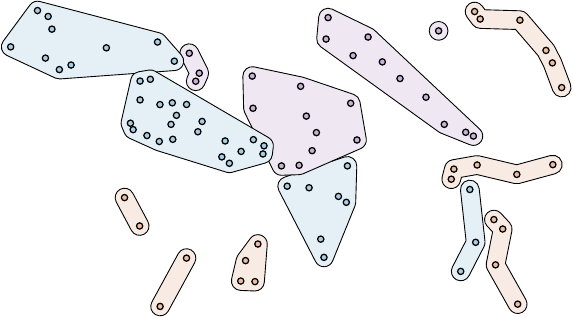}}
        \caption{Time $t = 4.5r_d$}
    \end{subfigure}
    \hfill
    \begin{subfigure}{\columnwidth}
        \fbox{\includegraphics[width=0.97\columnwidth]{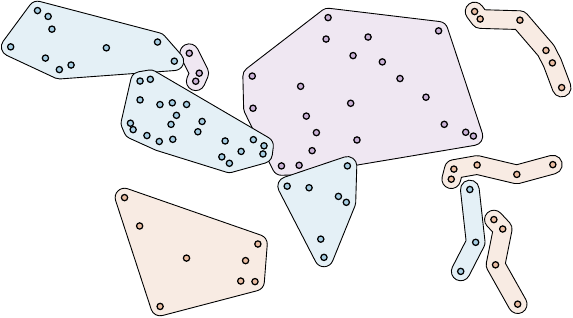}}
        \caption{Time $t = 6r_d$}
    \end{subfigure}
    \caption{SimpleSets visualizations of the Hotels dataset~\cite{LineSets, KelpFusion} based on different partitions. Eleven points of the original dataset were removed because they belong to multiple sets.}
    \label{app:fig:times}
\end{figure*}

\begin{figure*}[b]
    \centering
    \begin{subfigure}{0.45\textwidth}
        \fbox{\includegraphics[width=0.96\textwidth]{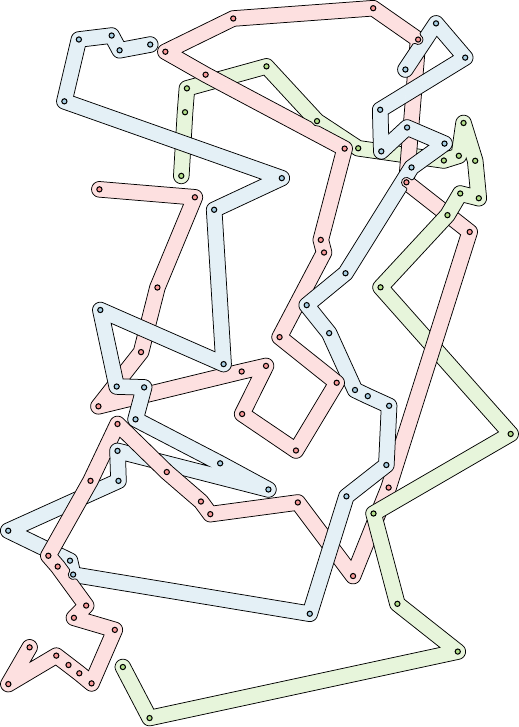}}
        \caption{LineSets}
        \label{app:fig:nyc-SimpleSets:LineSets}
    \end{subfigure}

    \vspace{2mm}

    \centering
    \begin{subfigure}{0.46\textwidth}
        \centering
        \fbox{\includegraphics[width=0.93\textwidth]{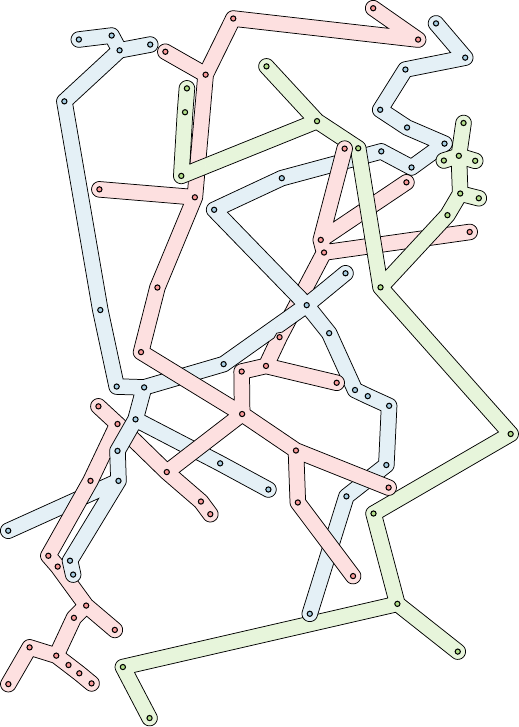}}
        \caption{Bubble Sets}
        \label{app:fig:nyc-SimpleSets:BubbleSets}
    \end{subfigure}
    \begin{subfigure}{0.46\textwidth}
        \centering
        \fbox{\includegraphics[width=0.93\textwidth]{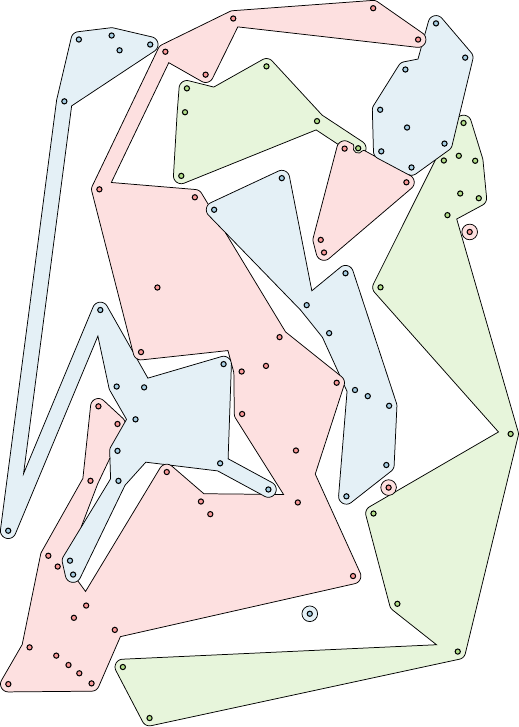}}
        \caption{ClusterSets}
        \label{app:fig:nyc-SimpleSets:ClusterSets}
    \end{subfigure}
    \caption{Patterns from other visualizations drawn with the SimpleSets drawing algorithm.}
    \label{app:fig:nyc-SimpleSets}
\end{figure*}

\begin{figure*}
    \centering
    \includegraphics[]{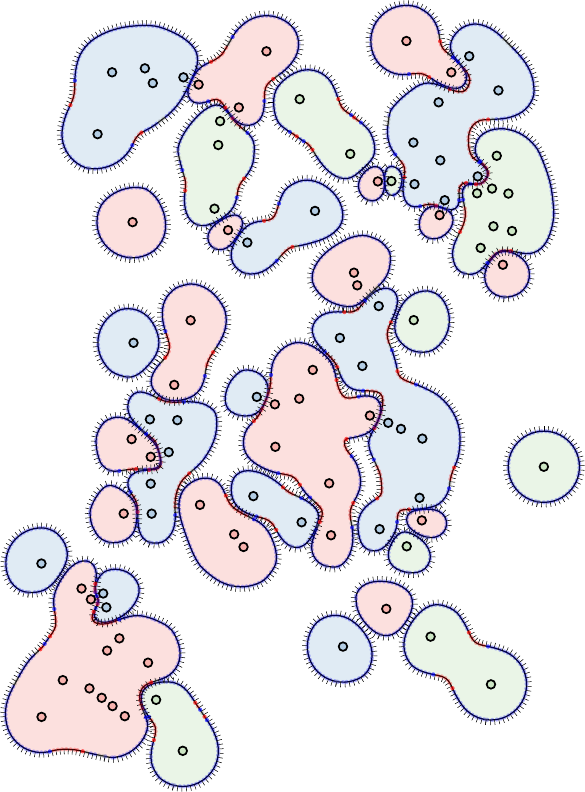}
    \caption{Equidistant point sampling used to measure inflections and curvature. The figure shows VPF*14+ on the NYC dataset. The black line segments show normals. The circles on the boundary show turns: clockwise (blue), counter-clockwise (red), and straight (grey).}
    \label{app:fig:sampling}
\end{figure*}

Figure~\ref{app:fig:times} 
shows four SimpleSets partitions for different time parameters. 
Figure~\ref{app:fig:nyc-SimpleSets} 
illustrates that any set of disjoint patterns can be drawn with the SimpleSets drawing algorithm. 
Figure~\ref{app:fig:sampling}
shows the equidistant point sampling used to measure inflections and curvature for the quantitative analysis.
The remaining figures are outputs not present in the paper that were analyzed in the quantitative evaluation.

\putbib

\clearpage

\clearpage

\begin{figure*}[tbh]
    \begin{subfigure}{\textwidth}
        \centering
        \frame{\includegraphics[width=0.6\textwidth, page=2]{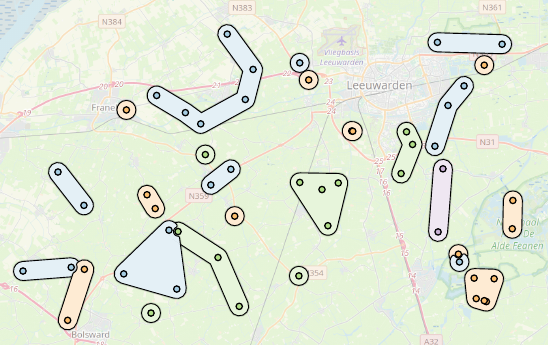}}
        \caption{SimpleSets}
        \label{app:fig:mills:SimpleSets}
    \end{subfigure}
    
    \vspace{4mm}

    \begin{subfigure}{0.49\textwidth}
        \frame{\includegraphics[width=\textwidth, page=3]{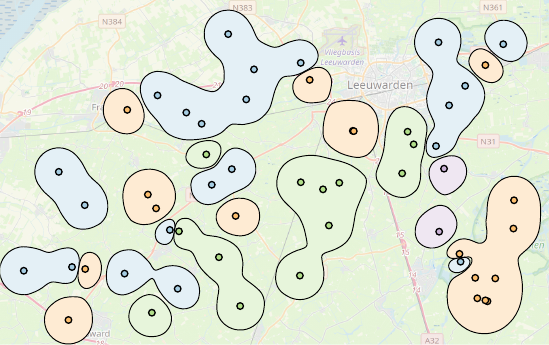}}
        \caption{VPF*14+}
        \label{app:fig:mills:VPF*14+}
    \end{subfigure}
    \hfill
    \begin{subfigure}{0.49\textwidth}
        \frame{\includegraphics[width=\textwidth, page=4]{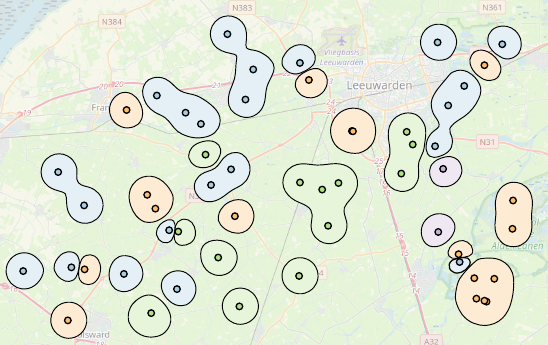}}
        \caption{VPF*14-}
        \label{app:fig:mills:VPF*14-}
    \end{subfigure}
    
    \vspace{4mm}
    
    \begin{subfigure}{\textwidth}
        \centering
        \frame{\includegraphics[width=0.6\textwidth, page=5]{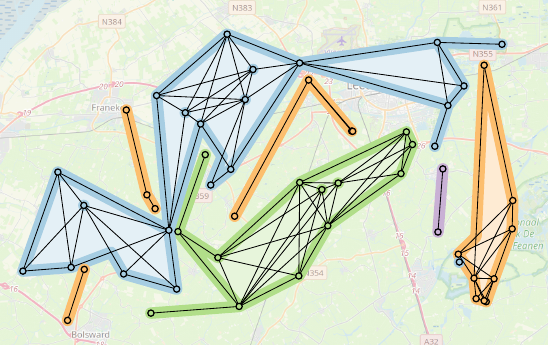}}
        \caption{ClusterSets}
        \label{app:fig:mills:ClusterSets}
    \end{subfigure}
    \caption{Comparison on the Mills dataset. The ClusterSets and VPF*14 outputs come from our implementation of their method.}
    \label{app:fig:mills}
\end{figure*}

\begin{figure*}
    \centering
    \frame{\includegraphics[width=0.5\textwidth, page=8]{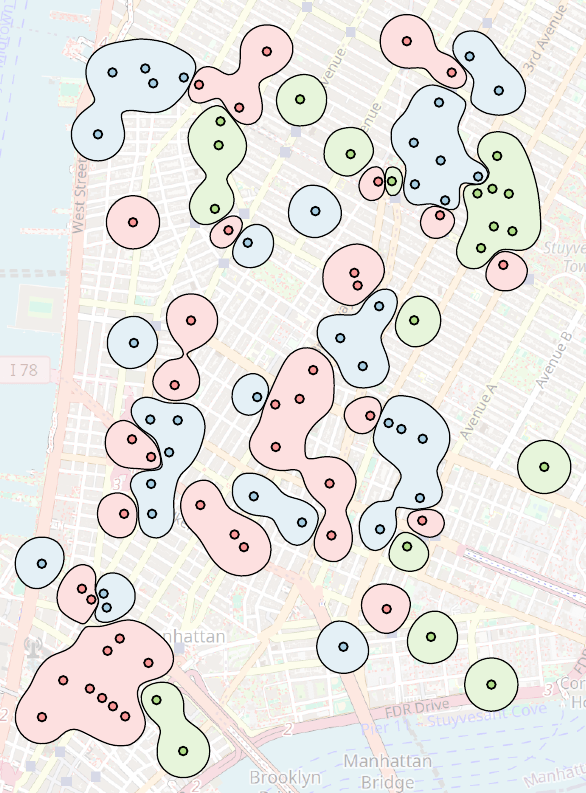}}
    \caption{VPF*14 visualization on the NYC dataset with medium radius of influence.}
    \label{app:fig:nyc:VPF*14-}
\end{figure*}

\begin{figure*}
    \includegraphics[width=\textwidth, page=1]{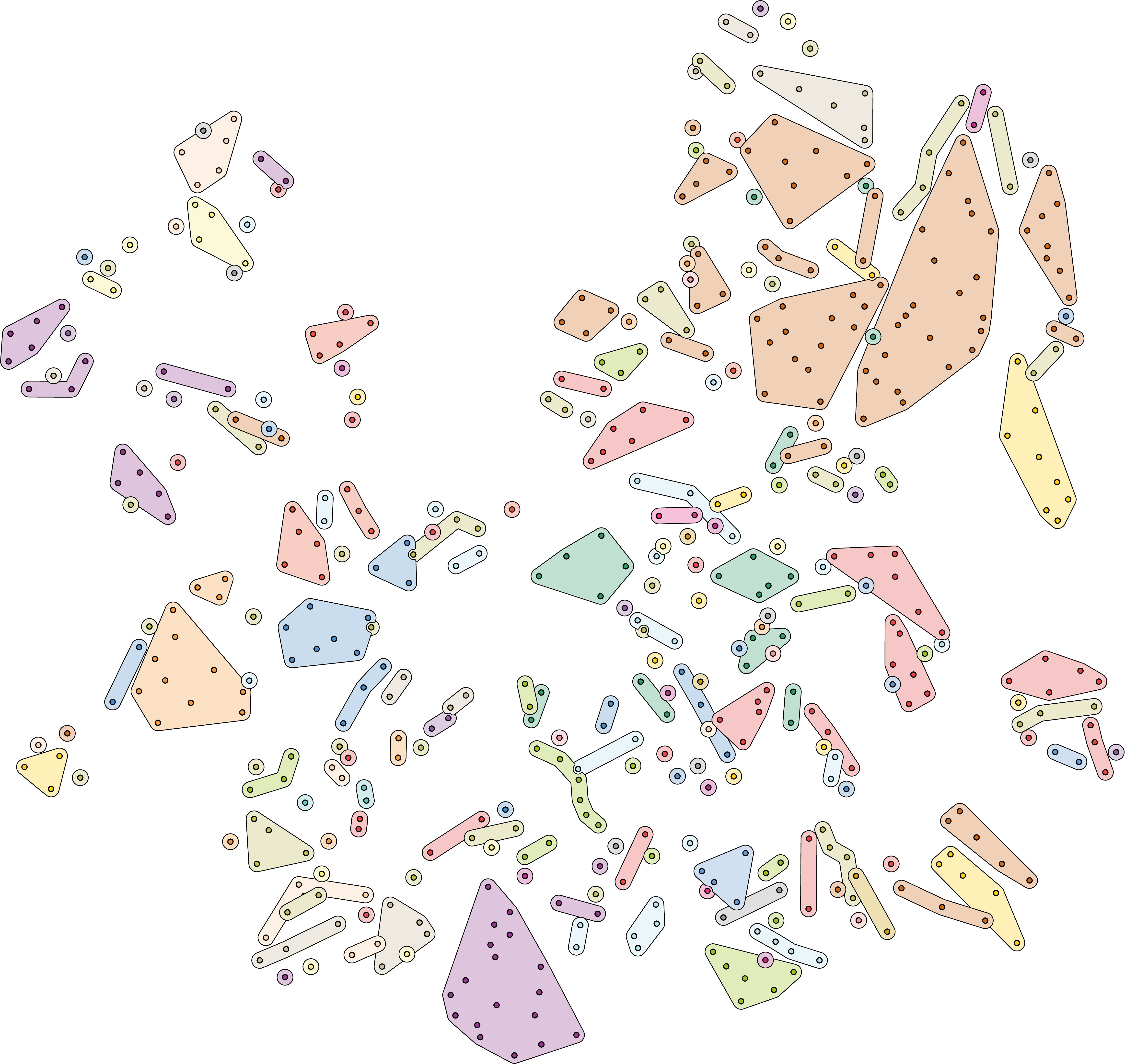}
    \caption{SimpleSets visualization of the human disease network.}
    \label{app:fig:diseasome:SimpleSets}
\end{figure*}

\begin{figure*}
    \includegraphics[width=\textwidth]{figures/Vihrovs-diseasome-full-new.pdf}
    \caption{VPF*14 visualization of the human disease network; generated using our implementation of their method. Large radius of influence.}
    \label{app:fig:diseasome:VPF*14+}
\end{figure*}

\begin{figure*}
    \includegraphics[width=\textwidth]{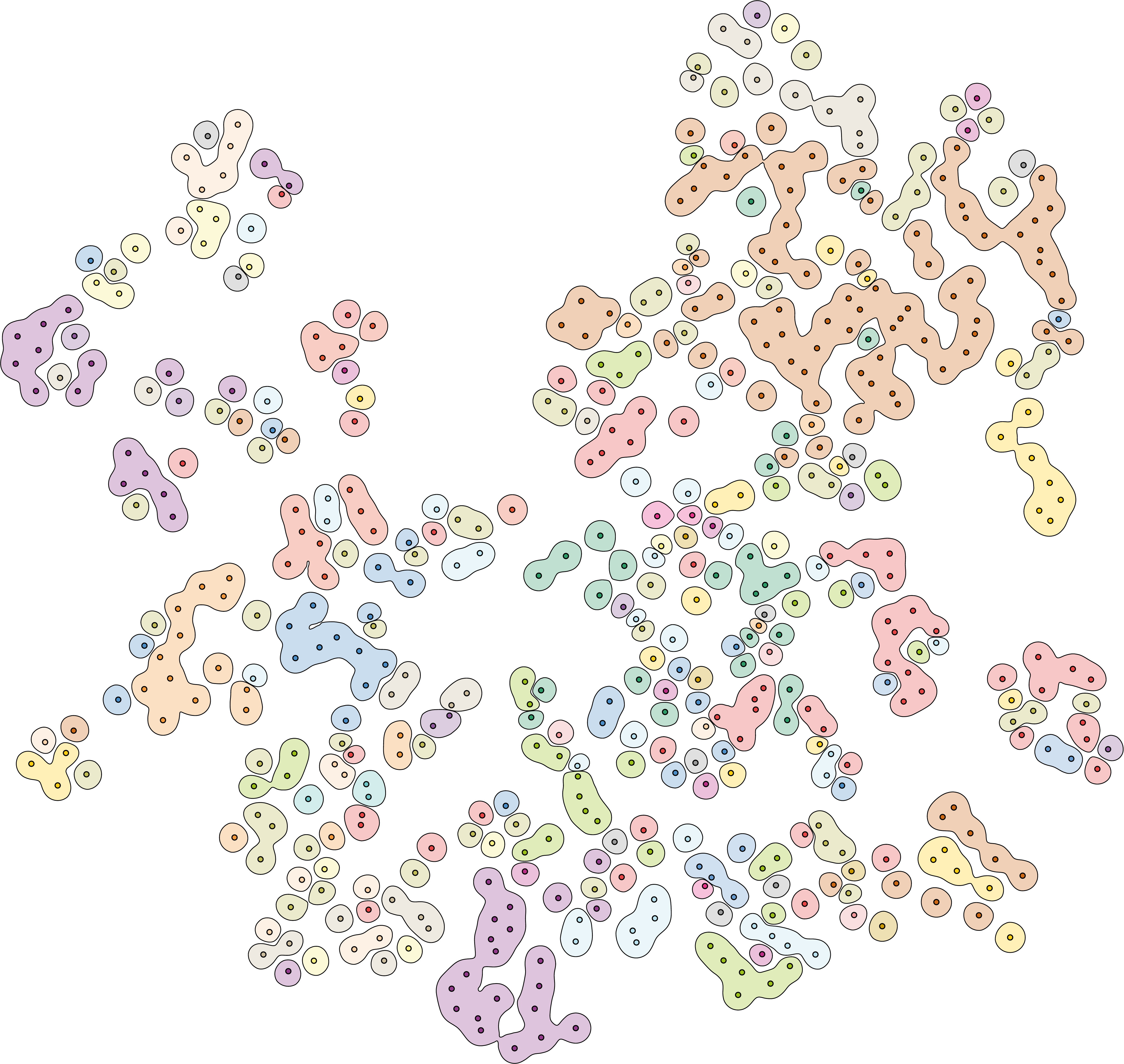}
    \caption{VPF*14 visualization of the human disease network; generated using our implementation of their method. Medium radius of influence.}
    \label{app:fig:diseasome:VPF*14-}
\end{figure*}

\begin{figure*}
    \includegraphics[width=\textwidth]{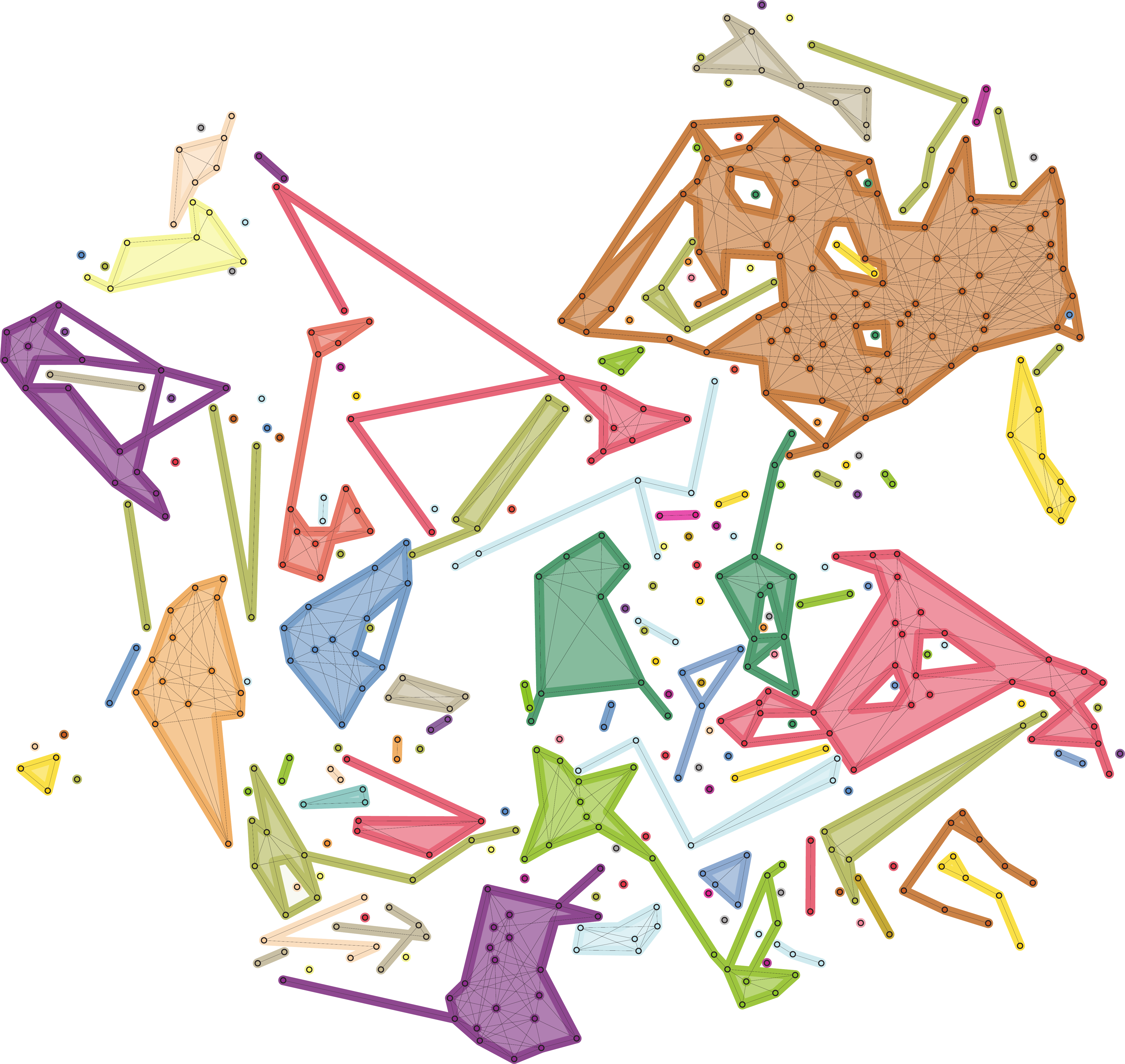}
    \caption{ClusterSets visualization of the human disease network; generated using our implementation of their method.}
    \label{app:fig:diseasome:ClusterSets}
\end{figure*}

\end{bibunit}

%% file: SimpleSets.bbl
blank

\begin{thebibliography}{10}

\bibitem{multi-colored-spanning-graphs}
H.~A. Akitaya, M.~L{\"{o}}ffler, and C.~D. T{\'{o}}th.
\newblock Multi-colored spanning graphs.
\newblock {\em Theoretical Computer Science}, 833:11--25, 2020.
  \href{https://doi.org/10.1016/J.TCS.2020.04.022}
{doi: {{%
10\hspace{.1pt}\discretionary{.}{%
}{.}\hspace{.4pt}1016\discretionary{/}{%
}{/}J\hspace{.1pt}\discretionary{.}{%
}{.}\hspace{.4pt}TCS\hspace{.1pt}\discretionary{.}{%
}{.}\hspace{.4pt}2020\hspace{.1pt}\discretionary{.}{%
}{.}\hspace{.4pt}04\hspace{.1pt}\discretionary{.}{%
}{.}\hspace{.4pt}022}}}


\bibitem{LineSets}
B.~Alper, N.~{Henry Riche}, G.~A. Ramos, and M.~Czerwinski.
\newblock Design study of {L}ine{S}ets, a novel set visualization technique.
\newblock {\em {IEEE} Transactions on Visualization and Computer Graphics},
  17(12):2259--2267, 2011. \href{https://doi.org/10.1109/TVCG.2011.186}
{doi: {{%
10\hspace{.1pt}\discretionary{.}{%
}{.}\hspace{.4pt}1109\discretionary{/}{%
}{/}TVCG\hspace{.1pt}\discretionary{.}{%
}{.}\hspace{.4pt}2011\hspace{.1pt}\discretionary{.}{%
}{.}\hspace{.4pt}186}}}


\bibitem{set-survey}
B.~Alsallakh, L.~Micallef, W.~Aigner, H.~Hauser, S.~Miksch, and P.~J. Rodgers.
\newblock The state-of-the-art of set visualization.
\newblock {\em Computer Graphics Forum}, 35(1):234--260, 2016.
  \href{https://doi.org/10.1111/CGF.12722}
{doi: {{%
10\hspace{.1pt}\discretionary{.}{%
}{.}\hspace{.4pt}1111\discretionary{/}{%
}{/}CGF\hspace{.1pt}\discretionary{.}{%
}{.}\hspace{.4pt}12722}}}


\bibitem{attneave1957physical}
F.~Attneave.
\newblock Physical determinants of the judged complexity of shapes.
\newblock {\em Journal of Experimental Psychology}, 53(4):221, 1957.
  \href{https://doi.org/10.1037/h0043921}
{doi: {{%
10\hspace{.1pt}\discretionary{.}{%
}{.}\hspace{.4pt}1037\discretionary{/}{%
}{/}h0043921}}}


\bibitem{optimal-islands}
C.~Bautista{-}Santiago, J.~M. D{\'{i}}az{-}B{\'{a}}{\~{n}}ez, D.~Lara,
  P.~P{\'{e}}rez{-}Lantero, J.~Urrutia, and I.~Ventura.
\newblock Computing optimal islands.
\newblock {\em Operations Research Letters}, 39(4):246--251, 2011.
  \href{https://doi.org/10.1016/j.orl.2011.04.008}
{doi: {{%
10\hspace{.1pt}\discretionary{.}{%
}{.}\hspace{.4pt}1016\discretionary{/}{%
}{/}j\hspace{.1pt}\discretionary{.}{%
}{.}\hspace{.4pt}orl\hspace{.1pt}\discretionary{.}{%
}{.}\hspace{.4pt}2011\hspace{.1pt}\discretionary{.}{%
}{.}\hspace{.4pt}04\hspace{.1pt}\discretionary{.}{%
}{.}\hspace{.4pt}008}}}


\bibitem{computational-geometry-book}
M.~de~Berg, O.~Cheong, M.~J. van Kreveld, and M.~H. Overmars.
\newblock {\em Computational geometry: algorithms and applications}.
\newblock Springer, third ed., 2008.
  \href{https://doi.org/10.1007/978-3-540-77974-2}
{doi: {{%
10\hspace{.1pt}\discretionary{.}{%
}{.}\hspace{.4pt}1007\discretionary{/}{%
}{/}978\discretionary{%
}{-}{-}3\discretionary{%
}{-}{-}540\discretionary{%
}{-}{-}77974\discretionary{%
}{-}{-}2}}}


\bibitem{DBLP:conf/gis/BrinkhoffKSB95}
T.~Brinkhoff, H.~Kriegel, R.~Schneider, and A.~Braun.
\newblock Measuring the complexity of polygonal objects.
\newblock In {\em Proc. 3rd {ACM} International Workshop on Advances in
  Geographic Information Systems}, p. 109. {ACM}, 1995.

\bibitem{AOI}
H.~Byelas and A.~C. Telea.
\newblock Towards realism in drawing areas of interest on architecture
  diagrams.
\newblock {\em Journal of Visual Languages \& Computing}, 20(2):110--128, 2009.
  \href{https://doi.org/10.1016/j.jvlc.2008.09.001}
{doi: {{%
10\hspace{.1pt}\discretionary{.}{%
}{.}\hspace{.4pt}1016\discretionary{/}{%
}{/}j\hspace{.1pt}\discretionary{.}{%
}{.}\hspace{.4pt}jvlc\hspace{.1pt}\discretionary{.}{%
}{.}\hspace{.4pt}2008\hspace{.1pt}\discretionary{.}{%
}{.}\hspace{.4pt}09\hspace{.1pt}\discretionary{.}{%
}{.}\hspace{.4pt}001}}}


\bibitem{proportional-symbol-maps}
S.~Cabello, H.~J. Haverkort, M.~J. van Kreveld, and B.~Speckmann.
\newblock Algorithmic aspects of proportional symbol maps.
\newblock {\em Algorithmica}, 58(3):543--565, 2010.
  \href{https://doi.org/10.1007/s00453-009-9281-8}
{doi: {{%
10\hspace{.1pt}\discretionary{.}{%
}{.}\hspace{.4pt}1007\discretionary{/}{%
}{/}s00453\discretionary{%
}{-}{-}009\discretionary{%
}{-}{-}9281\discretionary{%
}{-}{-}8}}}


\bibitem{short-plane-supports}
T.~Castermans, M.~van Garderen, W.~Meulemans, M.~N{\"{o}}llenburg, and X.~Yuan.
\newblock Short plane supports for spatial hypergraphs.
\newblock {\em Journal of Graph Algorithms and Applications}, 23(3):463--498,
  2019. \href{https://doi.org/10.7155/JGAA.00499}
{doi: {{%
10\hspace{.1pt}\discretionary{.}{%
}{.}\hspace{.4pt}7155\discretionary{/}{%
}{/}JGAA\hspace{.1pt}\discretionary{.}{%
}{.}\hspace{.4pt}00499}}}


\bibitem{BubbleSets}
C.~Collins, G.~Penn, and S.~Carpendale.
\newblock Bubble {S}ets: Revealing set relations with isocontours over existing
  visualizations.
\newblock {\em {IEEE} Transactions on Visualization and Computer Graphics},
  15(6):1009--1016, 2009. \href{https://doi.org/10.1109/TVCG.2009.122}
{doi: {{%
10\hspace{.1pt}\discretionary{.}{%
}{.}\hspace{.4pt}1109\discretionary{/}{%
}{/}TVCG\hspace{.1pt}\discretionary{.}{%
}{.}\hspace{.4pt}2009\hspace{.1pt}\discretionary{.}{%
}{.}\hspace{.4pt}122}}}


\bibitem{DBLP:journals/vc/Dai0ZMLY22}
L.~Dai, K.~Zhang, X.~S. Zheng, R.~R. Martin, Y.~Li, and J.~Yu.
\newblock Visual complexity of shapes: a hierarchical perceptual learning
  model.
\newblock {\em Visual Computer}, 38(2):419--432, 2022.
  \href{https://doi.org/10.1007/S00371-020-02023-Z}
{doi: {{%
10\hspace{.1pt}\discretionary{.}{%
}{.}\hspace{.4pt}1007\discretionary{/}{%
}{/}S00371\discretionary{%
}{-}{-}020\discretionary{%
}{-}{-}02023\discretionary{%
}{-}{-}Z}}}


\bibitem{kelp-diagrams}
K.~Dinkla, M.~J. van Kreveld, B.~Speckmann, and M.~A. Westenberg.
\newblock Kelp {D}iagrams: Point set membership visualization.
\newblock {\em Computer Graphics Forum}, 31(3pt1):875--884, 2012.
  \href{https://doi.org/10.1111/j.1467-8659.2012.03080.x}
{doi: {{%
10\hspace{.1pt}\discretionary{.}{%
}{.}\hspace{.4pt}1111\discretionary{/}{%
}{/}j\hspace{.1pt}\discretionary{.}{%
}{.}\hspace{.4pt}1467\discretionary{%
}{-}{-}8659\hspace{.1pt}\discretionary{.}{%
}{.}\hspace{.4pt}2012\hspace{.1pt}\discretionary{.}{%
}{.}\hspace{.4pt}03080\hspace{.1pt}\discretionary{.}{%
}{.}\hspace{.4pt}x}}}


\bibitem{monochromatic-parts}
A.~Dumitrescu and J.~Pach.
\newblock Partitioning colored point sets into monochromatic parts.
\newblock {\em International Journal of Computational Geometry \&
  Applications}, 12(5):401--412, 2002.
  \href{https://doi.org/10.1142/S0218195902000943}
{doi: {{%
10\hspace{.1pt}\discretionary{.}{%
}{.}\hspace{.4pt}1142\discretionary{/}{%
}{/}S0218195902000943}}}


\bibitem{MapSets}
A.~Efrat, Y.~Hu, S.~G. Kobourov, and S.~Pupyrev.
\newblock Map{S}ets: Visualizing embedded and clustered graphs.
\newblock {\em Journal of Graph Algorithms and Applications}, 19(2):571--593,
  2015. \href{https://doi.org/10.7155/jgaa.00364}
{doi: {{%
10\hspace{.1pt}\discretionary{.}{%
}{.}\hspace{.4pt}7155\discretionary{/}{%
}{/}jgaa\hspace{.1pt}\discretionary{.}{%
}{.}\hspace{.4pt}00364}}}


\bibitem{tunnels-bridges-switches}
D.~Eppstein, M.~J. van Kreveld, E.~Mumford, and B.~Speckmann.
\newblock Edges and switches, tunnels and bridges.
\newblock {\em Computational Geometry}, 42(8):790--802, 2009.
  \href{https://doi.org/10.1016/J.COMGEO.2008.05.005}
{doi: {{%
10\hspace{.1pt}\discretionary{.}{%
}{.}\hspace{.4pt}1016\discretionary{/}{%
}{/}J\hspace{.1pt}\discretionary{.}{%
}{.}\hspace{.4pt}COMGEO\hspace{.1pt}\discretionary{.}{%
}{.}\hspace{.4pt}2008\hspace{.1pt}\discretionary{.}{%
}{.}\hspace{.4pt}05\hspace{.1pt}\discretionary{.}{%
}{.}\hspace{.4pt}005}}}


\bibitem{GMap}
E.~R. Gansner, Y.~Hu, and S.~G. Kobourov.
\newblock {GM}ap: Visualizing graphs and clusters as maps.
\newblock In {\em Proc. Pacific Visualization Symposium (PacificVis)}, pp.
  201--208. {IEEE}, 2010.
  \href{https://doi.org/10.1109/PACIFICVIS.2010.5429590}
{doi: {{%
10\hspace{.1pt}\discretionary{.}{%
}{.}\hspace{.4pt}1109\discretionary{/}{%
}{/}PACIFICVIS\hspace{.1pt}\discretionary{.}{%
}{.}\hspace{.4pt}2010\hspace{.1pt}\discretionary{.}{%
}{.}\hspace{.4pt}5429590}}}


\bibitem{ClusterSets}
J.~Geiger, S.~Cornelsen, J.~Haunert, P.~Kindermann, T.~Mched\-lidze,
  M.~N{\"{o}}llenburg, Y.~Okamoto, and A.~Wolff.
\newblock Cluster{S}ets: Optimizing planar clusters in categorical point data.
\newblock {\em Computer Graphics Forum}, 40(3):471--481, 2021.
  \href{https://doi.org/10.1111/cgf.14322}
{doi: {{%
10\hspace{.1pt}\discretionary{.}{%
}{.}\hspace{.4pt}1111\discretionary{/}{%
}{/}cgf\hspace{.1pt}\discretionary{.}{%
}{.}\hspace{.4pt}14322}}}


\bibitem{human-disease-network}
K.-I. Goh, M.~E. Cusick, D.~Valle, B.~Childs, M.~Vidal, and A.-L. Barabási.
\newblock The human disease network.
\newblock {\em Proceedings of the National Academy of Sciences},
  104(21):8685--8690, 2007. \href{https://doi.org/10.1073/pnas.0701361104}
{doi: {{%
10\hspace{.1pt}\discretionary{.}{%
}{.}\hspace{.4pt}1073\discretionary{/}{%
}{/}pnas\hspace{.1pt}\discretionary{.}{%
}{.}\hspace{.4pt}0701361104}}}


\bibitem{social-networks-convex2}
J.~Heer and d.~boyd.
\newblock Vizster: Visualizing online social networks.
\newblock In {\em Proc. {IEEE} Symposium on Information Visualization
  (InfoVis)}, pp. 32--39. {IEEE}, 2005.
  \href{https://doi.org/10.1109/INFVIS.2005.1532126}
{doi: {{%
10\hspace{.1pt}\discretionary{.}{%
}{.}\hspace{.4pt}1109\discretionary{/}{%
}{/}INFVIS\hspace{.1pt}\discretionary{.}{%
}{.}\hspace{.4pt}2005\hspace{.1pt}\discretionary{.}{%
}{.}\hspace{.4pt}1532126}}}


\bibitem{untangling-euler-diagrams}
N.~{Henry Riche} and T.~Dwyer.
\newblock Untangling {E}uler diagrams.
\newblock {\em {IEEE} Transactions on Visualization and Computer Graphics},
  16(6):1090--1099, 2010. \href{https://doi.org/10.1109/TVCG.2010.210}
{doi: {{%
10\hspace{.1pt}\discretionary{.}{%
}{.}\hspace{.4pt}1109\discretionary{/}{%
}{/}TVCG\hspace{.1pt}\discretionary{.}{%
}{.}\hspace{.4pt}2010\hspace{.1pt}\discretionary{.}{%
}{.}\hspace{.4pt}210}}}


\bibitem{colored-spanning-graphs}
F.~Hurtado, M.~Korman, M.~J. van Kreveld, M.~L{\"{o}}ffler, V.~Sacrist{\'{a}}n,
  A.~Shioura, R.~I. Silveira, B.~Speckmann, and T.~Tokuyama.
\newblock Colored spanning graphs for set visualization.
\newblock {\em Computational Geometry}, 68:262--276, 2018.
  \href{https://doi.org/10.1016/J.COMGEO.2017.06.006}
{doi: {{%
10\hspace{.1pt}\discretionary{.}{%
}{.}\hspace{.4pt}1016\discretionary{/}{%
}{/}J\hspace{.1pt}\discretionary{.}{%
}{.}\hspace{.4pt}COMGEO\hspace{.1pt}\discretionary{.}{%
}{.}\hspace{.4pt}2017\hspace{.1pt}\discretionary{.}{%
}{.}\hspace{.4pt}06\hspace{.1pt}\discretionary{.}{%
}{.}\hspace{.4pt}006}}}


\bibitem{matsumoto2019quantification}
T.~Matsumoto, K.~Sato, Y.~Matsuoka, and T.~Kato.
\newblock Quantification of “complexity” in curved surface shape using
  total absolute curvature.
\newblock {\em Computers \& Graphics}, 78:108--115, 2019.

\bibitem{KelpFusion}
W.~Meulemans, N.~{Henry Riche}, B.~Speckmann, B.~Alper, and T.~Dwyer.
\newblock Kelp{F}usion: {A} hybrid set visualization technique.
\newblock {\em {IEEE} Transactions on Visualization and Computer Graphics},
  19(11):1846--1858, 2013. \href{https://doi.org/10.1109/TVCG.2013.76}
{doi: {{%
10\hspace{.1pt}\discretionary{.}{%
}{.}\hspace{.4pt}1109\discretionary{/}{%
}{/}TVCG\hspace{.1pt}\discretionary{.}{%
}{.}\hspace{.4pt}2013\hspace{.1pt}\discretionary{.}{%
}{.}\hspace{.4pt}76}}}


\bibitem{DBLP:journals/cgf/PaetzoldKSXSD23}
P.~Paetzold, R.~Kehlbeck, H.~Strobelt, Y.~Xue, S.~Storandt, and O.~Deussen.
\newblock {R}ect{E}uler: Visualizing intersecting sets using rectangles.
\newblock {\em Computer Graphics Forum}, 42(3):87--98, 2023.
  \href{https://doi.org/10.1111/CGF.14814}
{doi: {{%
10\hspace{.1pt}\discretionary{.}{%
}{.}\hspace{.4pt}1111\discretionary{/}{%
}{/}CGF\hspace{.1pt}\discretionary{.}{%
}{.}\hspace{.4pt}14814}}}


\bibitem{DBLP:conf/icip/PageKSRA03}
D.~L. Page, A.~F. Koschan, S.~R. Sukumar, B.~Roui{-}Abidi, and M.~A. Abidi.
\newblock Shape analysis algorithm based on information theory.
\newblock In {\em Proc. International Conference on Image Processing}, pp.
  229--232. {IEEE}, 2003. \href{https://doi.org/10.1109/ICIP.2003.1246940}
{doi: {{%
10\hspace{.1pt}\discretionary{.}{%
}{.}\hspace{.4pt}1109\discretionary{/}{%
}{/}ICIP\hspace{.1pt}\discretionary{.}{%
}{.}\hspace{.4pt}2003\hspace{.1pt}\discretionary{.}{%
}{.}\hspace{.4pt}1246940}}}


\bibitem{social-networks-convex1}
A.~Perer and B.~Shneiderman.
\newblock Balancing systematic and flexible exploration of social networks.
\newblock {\em {IEEE} Transactions on Visualization and Computer Graphics},
  12(5):693--700, 2006. \href{https://doi.org/10.1109/TVCG.2006.122}
{doi: {{%
10\hspace{.1pt}\discretionary{.}{%
}{.}\hspace{.4pt}1109\discretionary{/}{%
}{/}TVCG\hspace{.1pt}\discretionary{.}{%
}{.}\hspace{.4pt}2006\hspace{.1pt}\discretionary{.}{%
}{.}\hspace{.4pt}122}}}


\bibitem{DBLP:journals/comgeo/Rappaport91}
D.~Rappaport.
\newblock A convex hull algorithm for discs, and applications.
\newblock {\em Computational Geometry}, 1:171--187, 1991.
  \href{https://doi.org/10.1016/0925-7721(92)90015-K}
{doi: {{%
10\hspace{.1pt}\discretionary{.}{%
}{.}\hspace{.4pt}1016\discretionary{/}{%
}{/}0925\discretionary{%
}{-}{-}7721\discretionary{%
}{(}{(}92\discretionary{)}{%
}{)}90015\discretionary{%
}{-}{-}K}}}


\bibitem{tufte2001visual}
E.~R. Tufte.
\newblock {\em The visual display of quantitative information}.
\newblock Graphics Press, second ed., 2001.

\bibitem{inverse-distance}
J.~Vihrovs, K.~Prusis, K.~Freivalds, P.~Rucevskis, and V.~Krebs.
\newblock An inverse distance-based potential field function for overlapping
  point set visualization.
\newblock In {\em Proc. 5th International Conference on Information
  Visualization Theory and Applications}, pp. 29--38, 2014.
  \href{https://doi.org/10.5220/0004681100290038}
{doi: {{%
10\hspace{.1pt}\discretionary{.}{%
}{.}\hspace{.4pt}5220\discretionary{/}{%
}{/}0004681100290038}}}


\bibitem{gestalt}
J.~Wagemans, J.~H. Elder, M.~Kubovy, S.~E. Palmer, M.~A. Peterson, M.~Singh,
  and R.~von~der Heydt.
\newblock A century of gestalt psychology in visual perception: I. perceptual
  grouping and figure--ground organization.
\newblock {\em Psychological bulletin}, 138(6):1172, 2012.
  \href{https://doi.org/10.1037/a0029333}
{doi: {{%
10\hspace{.1pt}\discretionary{.}{%
}{.}\hspace{.4pt}1037\discretionary{/}{%
}{/}a0029333}}}


\bibitem{F2-Bubbles}
Y.~Wang, D.~Cheng, Z.~Wang, J.~Zhang, L.~Zhou, G.~He, and O.~Deussen.
\newblock F2-{B}ubbles: Faithful bubble set construction and flexible editing.
\newblock {\em {IEEE} Transactions on Visualization and Computer Graphics},
  28(1):422--432, 2022. \href{https://doi.org/10.1109/TVCG.2021.3114761}
{doi: {{%
10\hspace{.1pt}\discretionary{.}{%
}{.}\hspace{.4pt}1109\discretionary{/}{%
}{/}TVCG\hspace{.1pt}\discretionary{.}{%
}{.}\hspace{.4pt}2021\hspace{.1pt}\discretionary{.}{%
}{.}\hspace{.4pt}3114761}}}


\end{thebibliography}


\begin{thebibliography}{1}

\bibitem{LineSets}
B.~Alper, N.~{Henry Riche}, G.~A. Ramos, and M.~Czerwinski.
\newblock Design study of {L}ine{S}ets, a novel set visualization technique.
\newblock {\em {IEEE} Transactions on Visualization and Computer Graphics},
  17(12):2259--2267, 2011. \href{https://doi.org/10.1109/TVCG.2011.186}
{doi: {{%
10\hspace{.1pt}\discretionary{.}{%
}{.}\hspace{.4pt}1109\discretionary{/}{%
}{/}TVCG\hspace{.1pt}\discretionary{.}{%
}{.}\hspace{.4pt}2011\hspace{.1pt}\discretionary{.}{%
}{.}\hspace{.4pt}186}}}


\bibitem{DBLP:journals/pami/HaralickSZ87}
R.~M. Haralick, S.~R. Sternberg, and X.~Zhuang.
\newblock Image analysis using mathematical morphology.
\newblock {\em {IEEE} Transactions on Pattern Analysis and Machine
  Intelligence}, 9(4):532--550, 1987.
  \href{https://doi.org/10.1109/TPAMI.1987.4767941}
{doi: {{%
10\hspace{.1pt}\discretionary{.}{%
}{.}\hspace{.4pt}1109\discretionary{/}{%
}{/}TPAMI\hspace{.1pt}\discretionary{.}{%
}{.}\hspace{.4pt}1987\hspace{.1pt}\discretionary{.}{%
}{.}\hspace{.4pt}4767941}}}


\bibitem{KelpFusion}
W.~Meulemans, N.~{Henry Riche}, B.~Speckmann, B.~Alper, and T.~Dwyer.
\newblock Kelp{F}usion: {A} hybrid set visualization technique.
\newblock {\em {IEEE} Transactions on Visualization and Computer Graphics},
  19(11):1846--1858, 2013. \href{https://doi.org/10.1109/TVCG.2013.76}
{doi: {{%
10\hspace{.1pt}\discretionary{.}{%
}{.}\hspace{.4pt}1109\discretionary{/}{%
}{/}TVCG\hspace{.1pt}\discretionary{.}{%
}{.}\hspace{.4pt}2013\hspace{.1pt}\discretionary{.}{%
}{.}\hspace{.4pt}76}}}


\bibitem{DBLP:journals/comgeo/Rappaport91}
D.~Rappaport.
\newblock A convex hull algorithm for discs, and applications.
\newblock {\em Computational Geometry}, 1:171--187, 1991.
  \href{https://doi.org/10.1016/0925-7721(92)90015-K}
{doi: {{%
10\hspace{.1pt}\discretionary{.}{%
}{.}\hspace{.4pt}1016\discretionary{/}{%
}{/}0925\discretionary{%
}{-}{-}7721\discretionary{%
}{(}{(}92\discretionary{)}{%
}{)}90015\discretionary{%
}{-}{-}K}}}


\end{thebibliography}
